\documentclass[iop,useAMS,usenatbib,preprint]{emulateapj}
\usepackage{amsmath}
\usepackage{graphicx}
\usepackage[normalem]{ulem}

\defcitealias{BB2014}{Paper I}

\shortauthors{Bailey, Basu \& Caselli}

\begin{document}

\title{Kinematics in Partially Ionised Molecular Clouds: Implications for the Transition to Coherence}

\author{Nicole D. Bailey\altaffilmark{1,2}, Shantanu Basu\altaffilmark{3}, and Paola Caselli\altaffilmark{1,2}}
\altaffiltext{1}{Max-Planck-Institut f\"{u}r extraterrestrische Physik, Giessenbachstrasse 1, 85748 Garching, Germany}
\altaffiltext{2}{School of Physics and Astronomy, University of Leeds, Leeds, United Kingdom, LS2 9JT}
\altaffiltext{3}{Department of Physics and Astronomy, University of Western Ontario, 1151 Richmond Street, London, Ontario, Canada, N6A 3K7}

\email{ndbailey@mpe.mpg.de (NDB); basu@uwo.ca (SB); }
\email{caselli@mpe.mpg.de (PC)}

\begin{abstract}

\citet{BB2014} show analysis of density and mass-to-flux ratio maps for simulations with either an ionisation profile which takes into account photoionisation (step-like profile) or a cosmic ray only ionisation profile. We extend this study to analyse the effect of these ionisation profiles on velocity structures, kinematics, and synthetic spectra. Clump regions are found to occur at the convergence of two flows with a low velocity region and velocity direction transition occurring at the junction. Models with evident substructure show that core formation occurs on the periphery of these velocity valleys. Analysis of synthetic spectra reveals the presence of large non-thermal components within low-density gas, especially for models with the step-like ionisation profile. All cores show small, sub-thermal relative motions compared to background gas. Large deviations within this analysis are due to the line of sight intersecting low- and high-density regions across the velocity switch transition. Positive deviations correspond to a foreground core moving away from the observer while negative deviations correspond to a background core moving toward the observer. Comparison of velocities resulting from different ionisation profiles suggest that high ionisation fractions yield supersonic velocities, up to two times the sound speed, while regions with low ionisation fractions tend to be subsonic or mildly supersonic. This suggests that the transition to coherence within cores could be a transition between high and low ionisation fractions within the gas. 

\end{abstract}


\section{Introduction}

The formation, fragmentation and dynamical evolution of molecular clouds is still a matter of debate, with the astrophysical community divided among those favouring short evolutionary time scales set by turbulence \citep[e.g.][]{Hartmann2001, Hartmann2012} and those who take into account magnetic fields and their retarding effects in the dynamics \citep[e.g.][]{WM1997, Mouschovias2006}. The only way to set the debate is to compare results from theory and simulations inclusive of magnetic fields and non-ideal magnetohydrodynamic effects with detailed observations, preferably toward relatively simple and quiescent regions with no significant stellar feedback to reduce the number of unknowns in the models. 

Observers can probe internal motions and kinematics thanks to high spectral resolution observations of molecular tracers  \citep[e.g.][]{Goodman1993, Caselli2002a, Pineda2010Coherence, Hacar2013}. Such studies have shown that cloud cores in nearby low-mass star forming regions, are thermally supported, while turbulent non-thermal motions start to dominate in the sharp transition region between the core and the surrounding parent molecular cloud. Distant high-mass star forming regions, away from active sites of star formation, also show moderately supersonic motions when viewed with high spectral and angular resolution instruments \citep[e.g.][]{Henshaw2014}. Thus, spectroscopic observations are unique tools to study the dynamical evolution of molecular clouds and simulations need to reproduce these basic observational results. 

The choice of the right tracer to study the various parts of a molecular cloud is set by our understanding of astrochemical processes, which also play a crucial role on the evolution of the cloud. In fact, molecules such as CO and its isotopologues act as efficient cooling agents to allow a region to attain the temperatures required for star formation \citep[e.g][]{Goldsmith2001}. The ionisation fraction, set by an interplay of cosmic-ray ionisation and ion-molecule chemical reactions \citep[e.g.][]{Guelin1977, Caselli1998}, also has a profound effect on the evolution of a magnetised molecular cloud \citep[e.g][]{Shu1987}: high ionisation fractions within a magnetic region prevents the collapse of the cloud due to frequent collisions between neutrals and ions; conversely, at low ionisation fractions, neutrals can slip past the ions allowing for collapse to occur. As shown by \citet{BB2012}, the ionisation fraction within a cloud has a great effect on how the cloud will fragment. High ionisation not only increases the timescale for collapse, but also the lengthscale for fragmentation while low ionisation allows for smaller structures to from. Application of a step-like ionisation profile based on the results of photochemical studies by \citet{Ruffle1998} revealed a two-stage fragmentation process which can form subparsec cores within a $\sim$pc size clump.  Conversely, applying a cosmic ray only ionisation profile results in the formation of only subparsec cores \citep{BB2012}.

That being said, few simulations take into account the specific chemistry of a molecular cloud when examining its evolution and the formation of stars. \citet[][hereafter Paper I]{BB2014}, presented the results of non-ideal magnetohydrodynamic (MHD) simulations of the two-stage fragmentation model. Specifically, \citetalias{BB2014} explored the effects of a step-like ionisation profile and microturbulence via ongoing density perturbations on core collapse in an effort to determine the necessary parameters for the two-stage fragmentation process to occur. To that end, they only presented the density and mass-to-flux ratio ($\mu$) results and the physical parameters of the clumps and cores formed. Although the application of these ionisation profiles are by no means a full treatment of the complex chemistry within a molecular cloud, use of such a profile represents the first step to including the effect of the chemistry on the evolution of a molecular cloud. In this paper, we expand the analysis of these simulations to explore effects of the ionisation profile on the velocity structure and subsequent core formation within molecular clouds. As with \citetalias{BB2014}, these simulations will assume microturbulent perturbations. Here, we focus on the effects of the ionisation profile on the velocity structure and molecular line profiles. A following paper will expand this analysis to fully turbulent models. 

The rest of the paper is set up as follows. Section 2 describes the numerical code and models focusing on the details pertinent to the analysis goals of this paper. Section 3 looks at the velocity structure within the clumps/cores of each model. Section 4 presents analysis of synthetic spectra and the effects of the ionisation profiles on the velocity dispersion and centroid velocities. Finally Sections 5 and 6 discuss and summarise the results and trends revealed in the previous sections.

\section{Simulations}
\subsection{Numerical Code}
We explore the kinematics and velocity structures within clump-core complexes in partially ionised, isothermal magnetic interstellar molecular clouds. The simulations were performed using the non-ideal MHD IDL code developed by \citet{BC2004} and including the additional ionisation profiles described in \citetalias{BB2014}. This code assumes planar clouds with infinite extent in the $x$- and $y$ directions and a local vertical half thickness Z. A full description of the assumptions, nonaxisymmetric equations and formulations can be found in \citet{BC2004, CB2006, Basu2009a, Basu2009b}, however for convenience, we highlight those essential for the analysis within this paper. 

The model assumes a magnetic field that threads the cloud perpendicular to the $xy$ plane and includes the effects of ambipolar diffusion. The timescale for collisions between neutral particles and ions is
\begin{equation} 
\tau_{ni} = 1.4 \left(\frac{m_i +m_{H_2}}{m_i} \right) \frac{1}{n_i\langle\sigma w\rangle_{iH_2}},
\label{tni} 
\end{equation} 
where $m_{i}$ is the ion mass, $m_{H_2}$ is the mass of molecular hydrogen, $n_{i}$ is the number density of ions, and $\langle\sigma w\rangle_{iH_2}$ is the neutral-ion collision rate. Assuming collisions between H$_{2}$ and HCO$^+$, the neutral-ion collision rate is $1.69\times 10^{-9}$ cm$^{3}$ s$^{-1}$ \citep{MM1973}. The factor of 1.4 in Equation~\ref{tni} accounts for neglecting the inertia of helium in calculating the slowing-down time of the neutrals by collisions with ions \citep{CB2006, MC1999} 

The threshold for collapse within a molecular cloud is regulated by the normalized mass-to-flux ratio of the background reference state,
\begin{equation}
\mu_{0} \equiv 2\pi G^{1/2}\frac{\sigma_{n,0}}{B_{\rm ref}},
\end{equation}
where $(2\pi G^{1/2})^{-1}$ is the critical mass-to-flux ratio for gravitational collapse in the adopted model \citep{CB2006}, $\sigma_{n,0}$ is the initial mass column density and $B_{\rm ref}$ is the constant, uniform magnetic field strength of the background reference state far away from the sheet. In the limit where $\tau_{ni} \rightarrow 0$, frequent collision between the neutral particles and ions couple the neutrals to the magnetic field, that is, the medium is flux frozen. Under these conditions, subcritical regions ($\mu_{0} < 1$) are supported by the magnetic field and only supercritical regions ($\mu_{0} > 1$) may collapse within a finite time frame. Non-zero values of $\tau_{ni}$ are inversely dependent on the ion number density and therefore on the degree of ionisation for a fixed neutral density.

Finally, the model is characterized by several dimensionless free parameters including a dimensionless form of the initial neutral-ion collision time ($\tau_{ni,0}/t_{0}~\equiv~2\pi G\sigma_{n,0}\tau_{ni,0}/c_{s}$) and a dimensionless external pressure ($\tilde{P}_{\rm ext} \equiv 2 P_{\rm ext}/\pi G \sigma^{2}_{n,0}$).  Here, $c_{s}~=~(k_{B} T/m_{n})^{1/2}$ is the isothermal sound speed; $k_{B}$ is the Boltzmann constant, $T$ is the temperature in Kelvin, and $m_{n}$ is the mean mass of a neutral particle ($m_{n}~=~2.33$ amu). We normalize column densities by $\sigma_{n,0}$, length scales by $L_{0}~=~c_{s}^{2}/2\pi G \sigma_{n,0}$ and time scales by $t_{0}~=~c_{s}/2\pi G \sigma_{n,0}$. Based on these parameters, typical values of the units used and other derived quantities are 

\begin{eqnarray}
\nonumber\sigma_{n,0} &=& \frac{3.63\times 10^{-3}}{(1+\tilde{P}_{\rm ext})^{1/2}}\left(\frac{n_{n,0}}{10^3 \rm ~cm^{-3}}\right)^{1/2}\left(\frac{T}{10 ~\rm K}\right)^{1/2} \rm g~cm^{-2},\\ 
&&\\
c_{s} &=& 0.188\left(\frac{T}{10 ~\rm K}\right)^{1/2} \rm km~s^{-1},\\
t_{0} &=& 3.98\times 10^5\left(\frac{10^3 \rm~ cm^{-3}}{n_{n,0}}\right)^{1/2}(1 + \tilde{P}_{\rm ext})^{1/2}~\rm yr,\label{time}\\ 
\nonumber L_{0} &=& 7.48\times 10^{-2} \left(\frac{T}{10 ~\rm K}\right)^{1/2}\times\\
&&\left(\frac{10^3 \rm ~cm^{-3}}{n_{n,0}}\right)^{1/2}(1 + \tilde{P}_{\rm ext})^{1/2}~\rm pc,\label{length}
\end{eqnarray}
where $n_{n,0}$ is the initial neutral number density. For our analysis, we assume a dimensionless external pressure $\tilde{P}_{\rm ext} = 0.1$ ($P_{\rm ext}/k_{B} \approx 10^{3}$ cm$^{-3}$ K) and a temperature $T = 10$ K. By assuming this value for $\tilde{P}_{ext}$, we are neglecting the effect of surface gravity waves, which act to reduce the fragmentation length and time scales \citep[see][]{CB2006,Basu2009b}. Our clouds are assumed to be evolving in isolation (i.e., not embedded within a cloud complex or adjacent to a hotter region) so that high external pressures would not be expected.

\subsection{Model Parameters}
\label{model}
The analysis presented in this paper expands upon that presented in \citetalias{BB2014}. As such the models presented here include four simulations previously presented within that paper and two extra models that were performed to round out the data set. As described in \citetalias{BB2014}, these simulations assume an initially diffuse cloud with an initial background column density which corresponds to a visual extinction $A_{V,0} = 1$ mag. Using the prescription of \citet{Pineda2010} \citep[see also][]{BB2012} and assuming a mean molecular weight of 2.33 amu, the resulting conversion between visual extinction and mass column density is 
\begin{equation} 
\sigma_{n} =3.638\times 10^{-3} (A_{V}/\rm mag)~\rm g~cm^{-2}.
\label{av2sigma}
\end{equation}

All simulations begin with an initial linear column density perturbation, $\delta\sigma_{n}/\sigma_{n,0}$ which is a normally distributed random variable with mean equal to zero and standard deviation $A$. The random value of $\delta\sigma_{n}/\sigma_{n,0}$ for each pixel is then added to the background column density in that pixel. To ensure that these perturbations do not artificially bias the simulation toward a particular fragmentation scale, all wavelengths are sampled and assigned to the region, i.e., they are white noise perturbations. For some simulations, 
\begin{figure*}
\centering
\includegraphics[width = 0.33\textwidth]{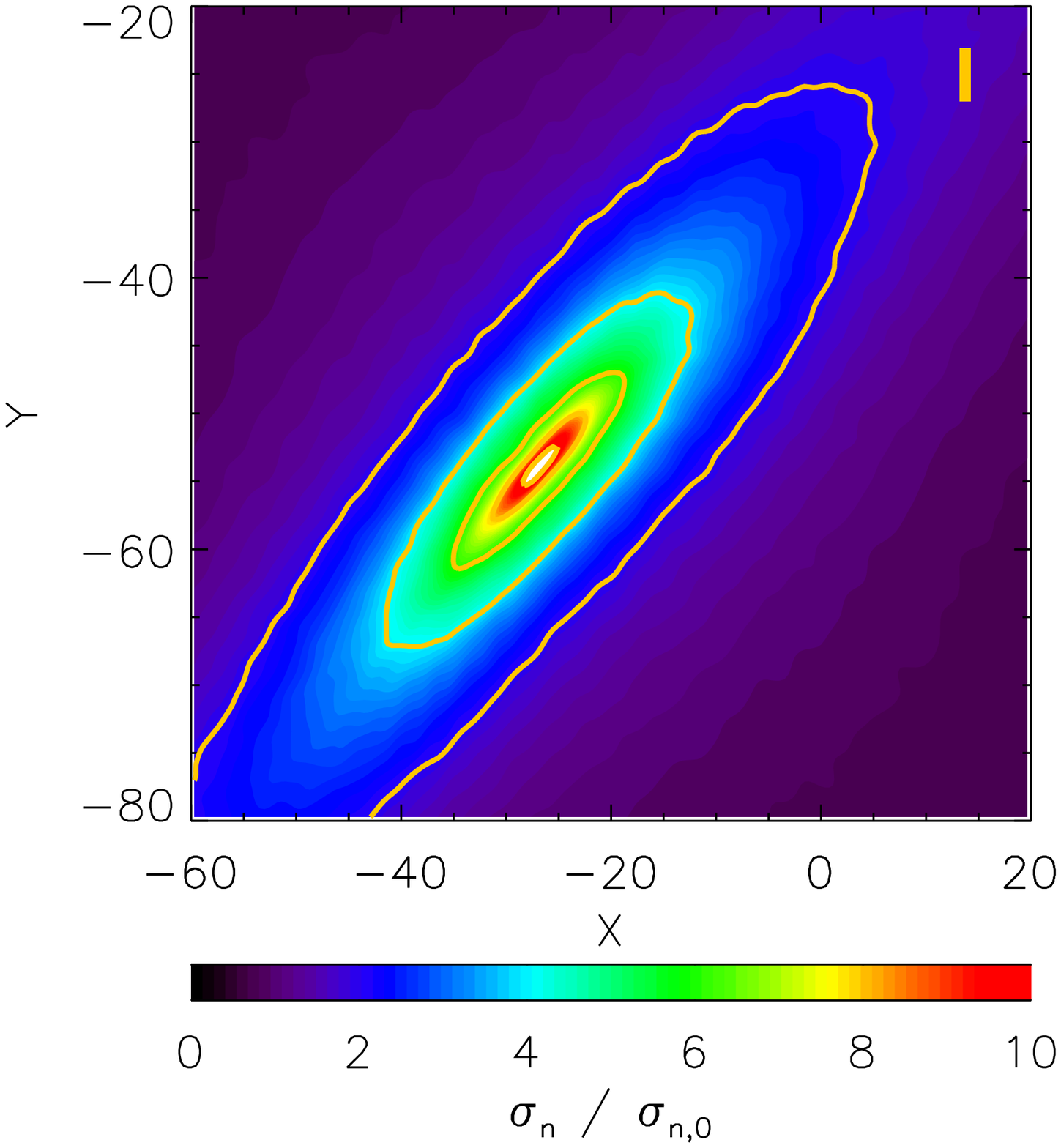}
\includegraphics[width = 0.33\textwidth]{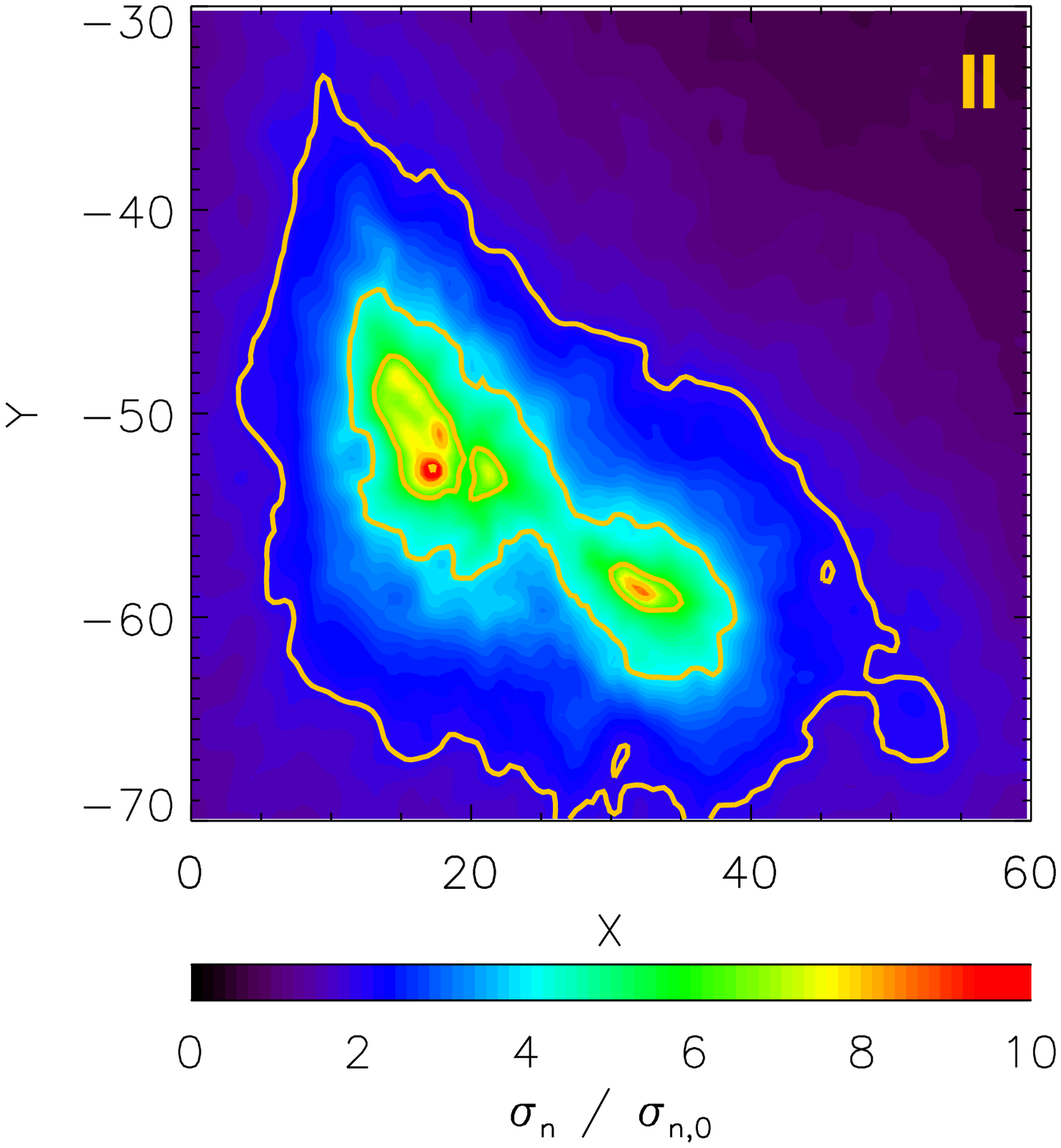}
\includegraphics[width = 0.33\textwidth]{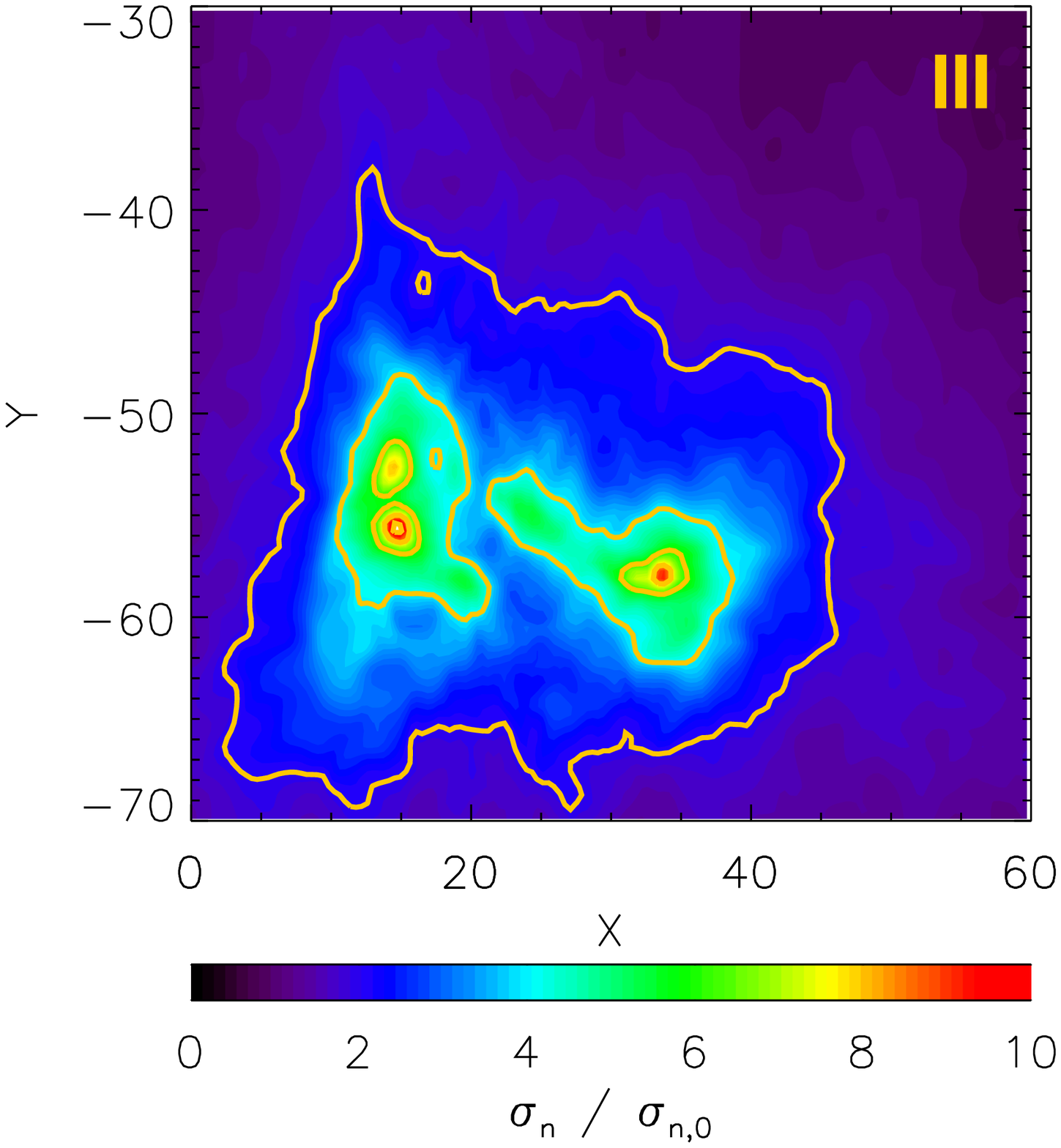}\\
\includegraphics[width = 0.33\textwidth]{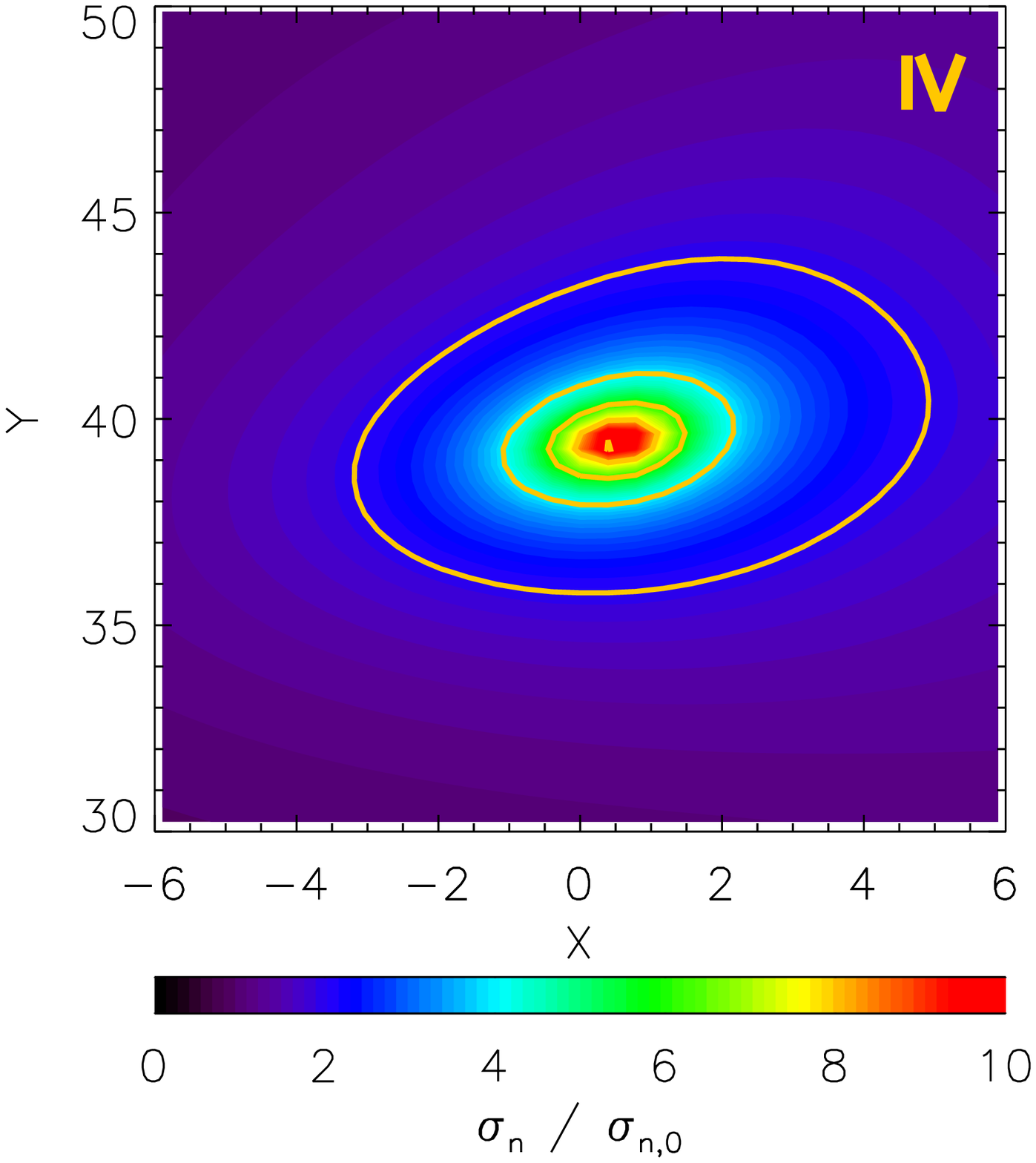}
\includegraphics[width = 0.33\textwidth]{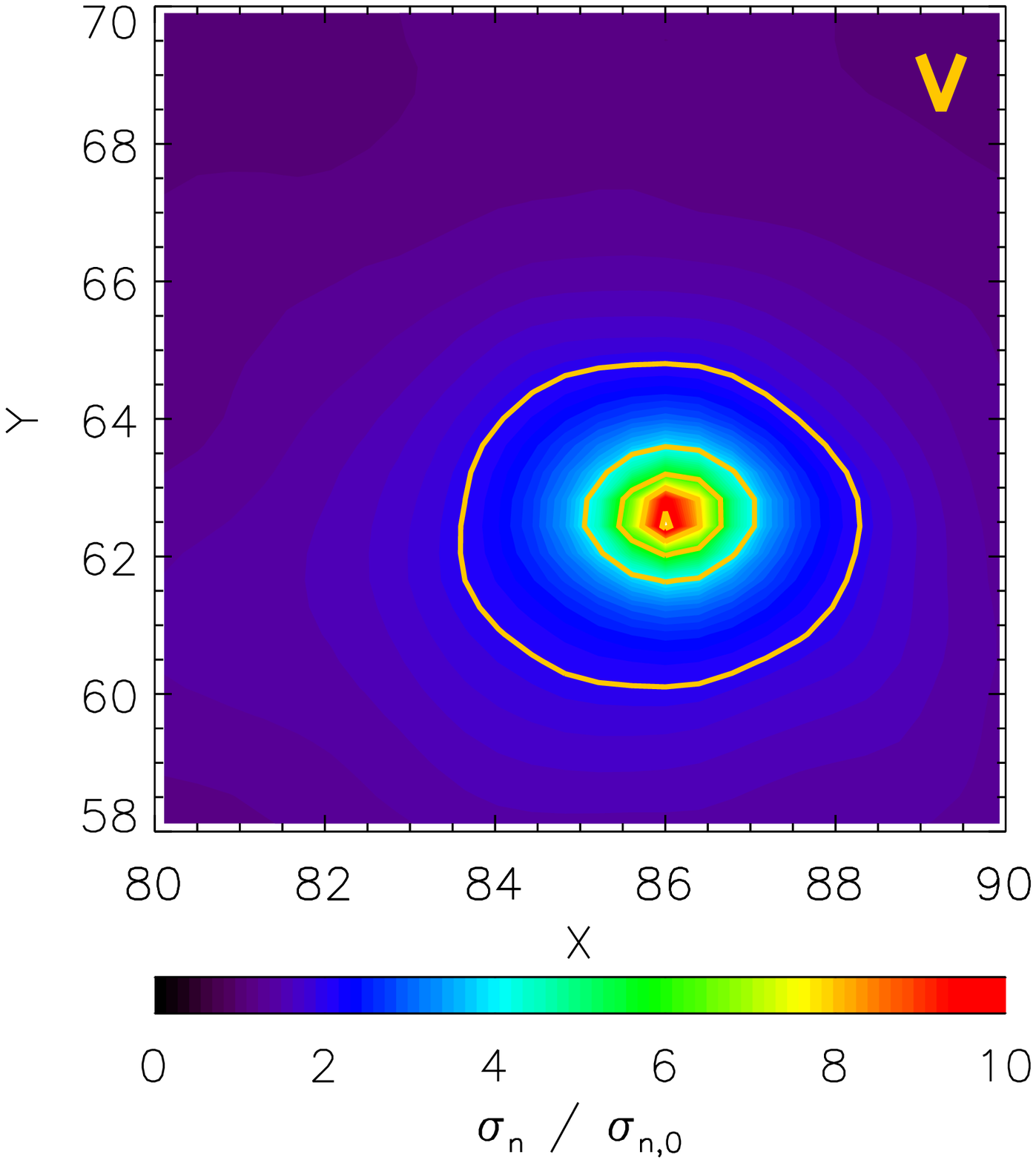}
\includegraphics[width = 0.33\textwidth]{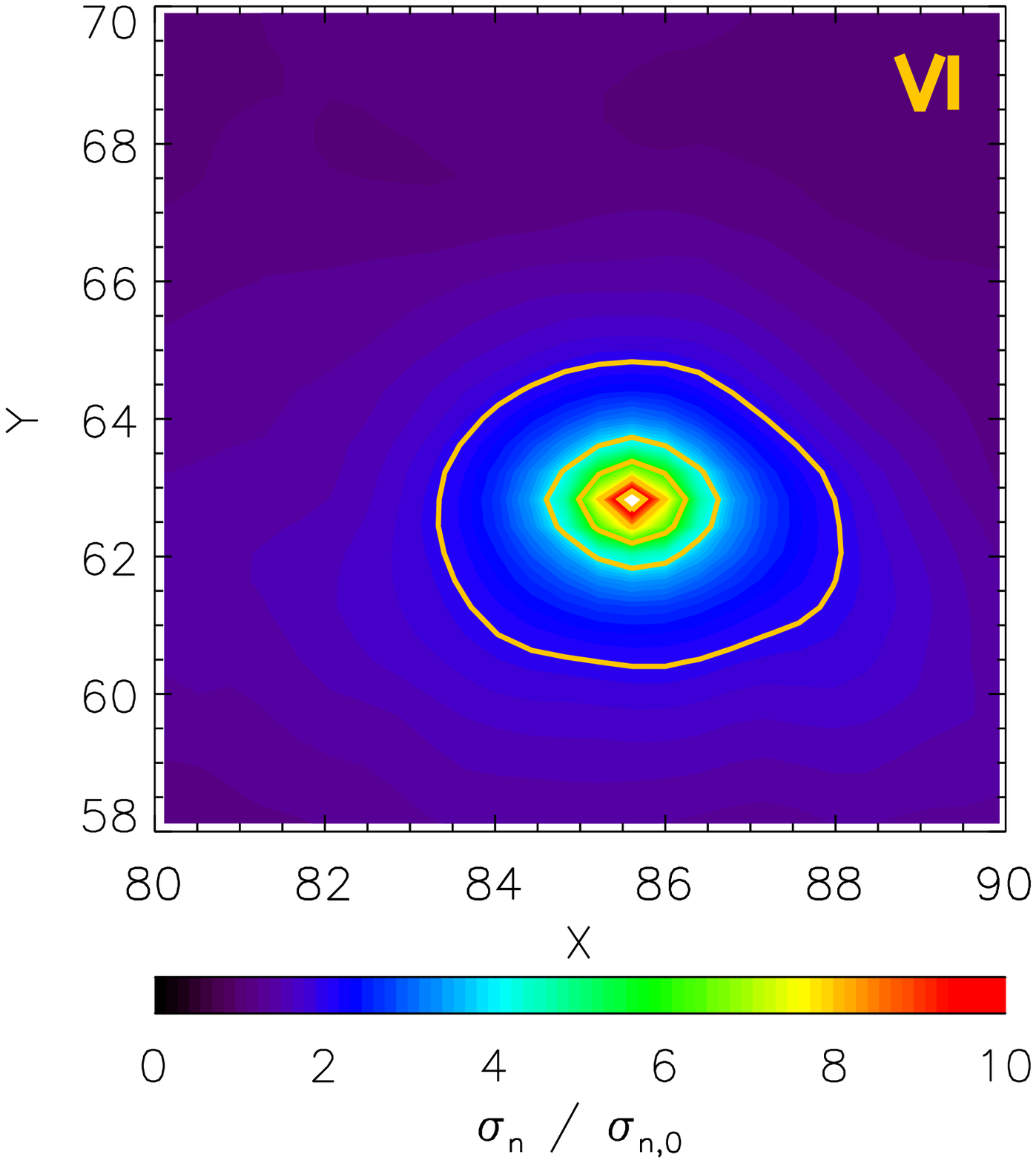}
\caption{Column density enhancement maps of clump/core regions. Each panel is a zoom-in of the full $64\pi L_{0}~x~64\pi L_{0}$ region that focuses on the region containing the most evolved clump/core in each model. Contours show the visual extinction value in 2 magnitude steps starting at $A_{V} = 2$. Panels are organized to depict increasing frequency of perturbations (initial perturbation only (left), perturbations every 10$t_{0}$ (middle), and perturbations every 5$t_{0}$ (right)) with top row depicting models with the step-like ionisation profile and the bottom row depicting models with the cosmic-ray only ionisation profile. Panels show each model at the last time of the respective simulations. Top row (from left to right): Model I ($t/t_{0} = 143.6$), Model II ($t/t_{0} = 80.7$), Model III ($t/t_{0} = 70.0$).  Bottom row (from left to right): Model IV ($t/t_{0} = 67.3$), Model V ($t/t_{0} = 40.9$), Model VI ($t/t_{0} = 35.4$). Axes are in units of $L_{0}$ (1 $L_{0}$ = 0.075 pc or 2.54 pixels.)}  
\label{sigmaoverlays}
\end{figure*}
subsequent perturbations are applied at specific intervals ($\Delta t_{sp}/t_{0}$). All simulations are performed on a 512 $\times$ 512 periodic box. The box size is 64$\pi L_{0}$, which is much larger than the preferred fragmentation scale ($4\pi L_{0}$) in the non-magnetic limit. For $T$ = 10~K and $\sigma_{n,0} = 3.638\times~10^{-3}$~g~cm$^{-2}$, $L_{0}~=~0.075$ pc. This translates to a box size of 15.16 pc or a pixel size of 0.0296 pc. For this analysis, all simulations assume perturbations with standard deviation about the mean of $A = 0.03$ and an initial mass-to-flux ratio $\mu_{0} = 1.1$. All other assumptions are the same as described in \citetalias{BB2014}. The initial parameters for the specific models can be found in Table~\ref{models}. The ionisation profiles quoted indicate the initial neutral-ion collision time. The SL profile (step-like) results in an initially almost flux frozen medium (i.e., $\tau_{ni,0}/t_{0} = 0.001$) while the CR profile results in a longer neutral-ion collision time ($\tau_{ni,0}/t_{0} = 0.2$). All simulations run until any pixel within the simulation equals or exceeds $\sigma_{n}/\sigma_{n,0} =10.$

\begin{deluxetable}{cccc}
\tablecaption{Simulation Parameters}
\tablewidth{0pt}
\tablehead{
\colhead{} & \colhead{\citetalias{BB2014}}  & \colhead{Ionisation} & \colhead{}\\
\colhead{Model} & \colhead{Model}  & \colhead{Profile\tablenotemark{*}} & \colhead{$\Delta t_{sp}/t_{0}$\tablenotemark{\dag}}
}
\startdata
I   & A & SL &  $\infty$  \\ 
II  & C & SL & 10        \\   
III & B & SL & 5       \\  
IV & -- & CR & $\infty$  \\  
V  & G & CR & 10        \\   
VI & --  & CR & 5  
\enddata
\tablenotetext{*}{Indicates the initial value of $\tau_{ni,0}/t_{0}$: $\tau_{ni,0}/t_{0}~=~0.001$ (SL) or  $\tau_{ni,0}/t_{0}~=~0.2$ (CR)\\
\dag~~$\Delta t_{sp}/t_{0}$ is the time between subsequent perturbations in dimensionless units. }
\label{models}
\end{deluxetable}

\subsection{Aims and Regions of Interest}
\label{aims}
Studies and observations \citep[][among others]{Kirk2009, Pineda2010Coherence, Walsh2004, Walsh2007} of star forming regions have found several properties regarding the kinematics of prestellar cores in relation to their surroundings. Specifically, cores are observed to have little internal turbulence \citep{BM1989, Jijina1999}, smaller velocity dispersions than the surrounding material \citep{BM1989, Goodman1998, Jijina1999, Pineda2010Coherence}, and small relative motions with respect to the surrounding material \citep{Walsh2004, Walsh2007}. \citet{Kirk2009} looked at the effect of turbulence and magnetic fields on the line widths of synthetic spectra created from thin disk simulations of molecular clouds. These simulations were performed using the same IDL MHD code as described above, however they only consider the CR ionisation profile. 

In this paper, we explore the effect of the assumed ionisation profile on the velocity field for simulations with ongoing column density perturbations. Specifically, we are interested in the regions of the simulation where clumps and/or cores have formed by the end of each run. From analysis of velocity maps within the plane, we look to determine how the ionisation profile shapes the velocity field. In addition, we create synthetic spectra assuming uniform optically thin conditions and constant fractional abundances of species representative of the cloud envelope (CO) and core ($N_{2}H^{+}$) to determine whether the cores formed within the simulations conform with the three kinematic properties of observed cores. 

\begin{figure}
\centering
\includegraphics[width = 0.48\textwidth]{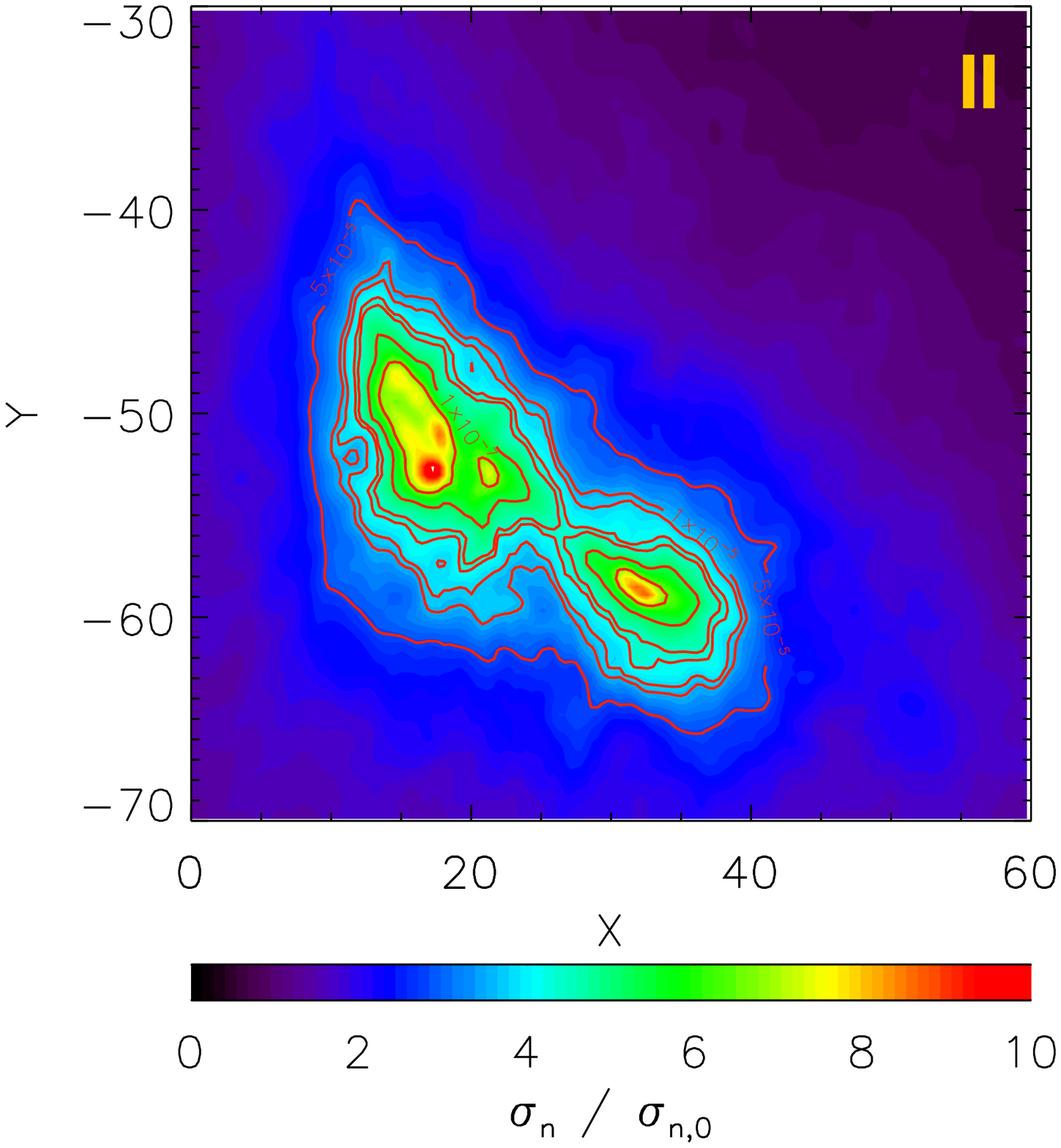}\\
\includegraphics[width = 0.48\textwidth]{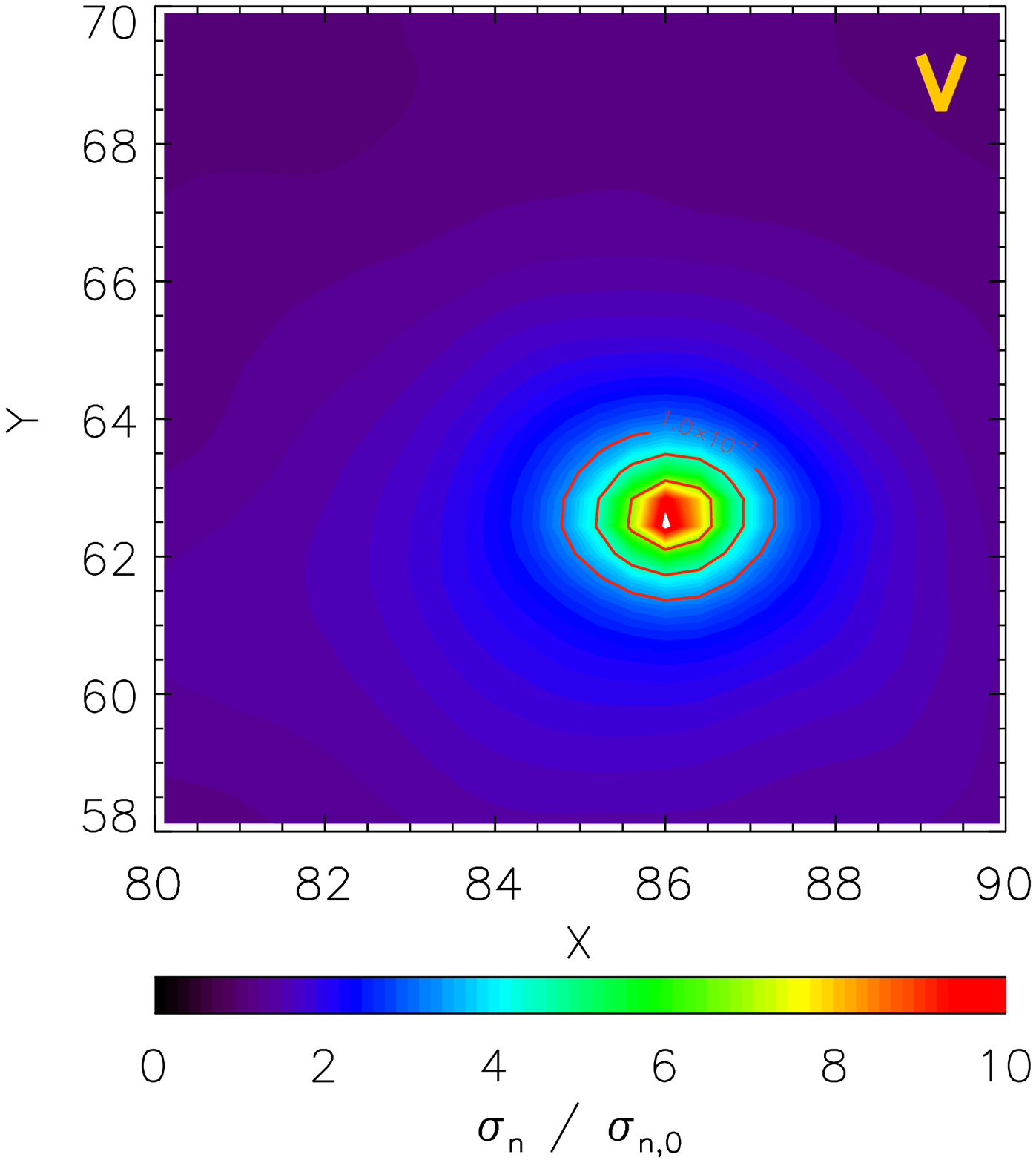}
\caption{Ionisation contours overlaid on column density enhancement maps for Model II (top) and Model V (bottom). Model II Contour levels: ($1.0\times 10^{-8}$, $5.0\times 10^{-8}$, $1.0\times 10^{-7}$, $5.0\times 10^{-7}$, $1.0\times 10^{-6}$, $5.0\times 10^{-6}$, $1.0\times 10^{-5}$, $5.0\times 10^{-5}$, $1.0\times 10^{-4}$). Model V Contour levels: ($5.0\times 10^{-8}$, $7.5\times 10^{-8}$, $1.0\times 10^{-7}$) (inner to outer respectively).}
\label{xi}
\end{figure}

Figure~\ref{sigmaoverlays} shows the density enhancement maps for each model. Each panel shows a zoom-in of the full simulation, focusing on the region containing the main clump or core within that model. The contours show the visual extinction in steps of 2 mag starting at $A_{V} = 2$ mag. Each panel is taken at the respective endpoint of each simulation (i.e., when one pixel within the full simulation reaches or exceeds $\sigma_{n}/\sigma_{n,0} = 10$) as indicated by the times in the caption. Figure~\ref{xi} shows the contours of the ionisation fraction overlaid on column density enhancement maps for two representative models: Model II (top) for the step-like ionisation profile and Model V (bottom) for the CR-only ionisation profile. The contour levels are indicated in the figure caption. As shown in the figure, the ionisation structure for both models follows the column density structure with the highest density regions exhibiting the lowest values of $\chi_{i}$. However there are some noticeable differences that are direct consequences of the profile shape. For Model V, we see that the ionisation contours are more or less evenly spaced, while for Model II, there is evidence of steep gradients in ionisation surrounding the various levels of structure within the clump including the clump itself. As expected, the entire core within Model V is encompassed by a low ionisation contour ($\chi_{i} = 1.0\times 10^{-7} $). In contrast, for Model II, the lowest ionisation contours outline the core envelopes while the clumps and low density gas are outlined by significantly higher ionisation contours. This is direct evidence of two stage fragmentation, where the larger fragmentation lengthscales occur at larger ionisation fractions and smaller lengthscales require smaller ionisation fractions \citepalias[see][]{BB2014}. 

\section{Velocity Structure within Molecular Clouds}

\begin{figure*}
\centering
\includegraphics[width = 0.33\textwidth]{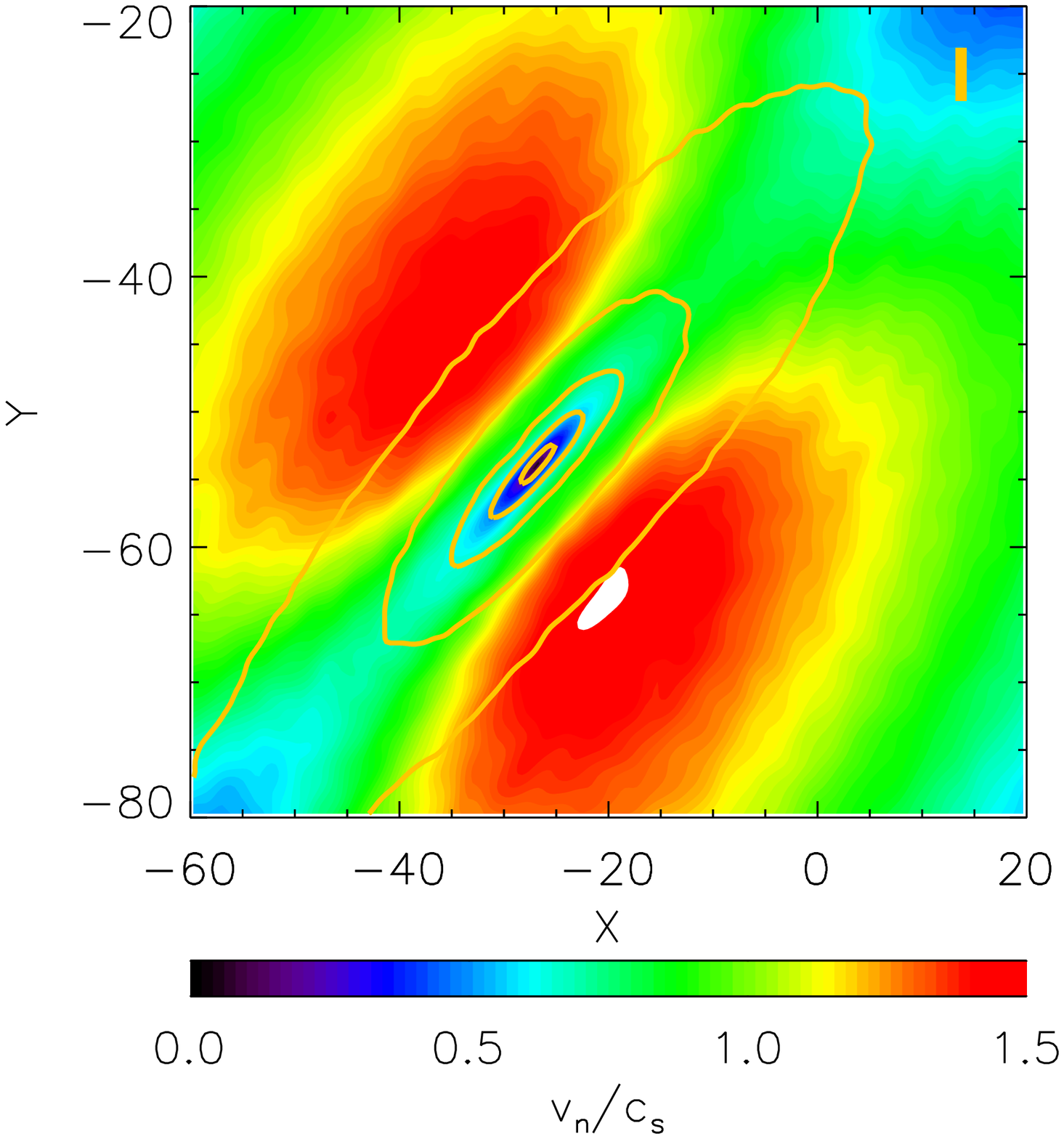}
\includegraphics[width = 0.33\textwidth]{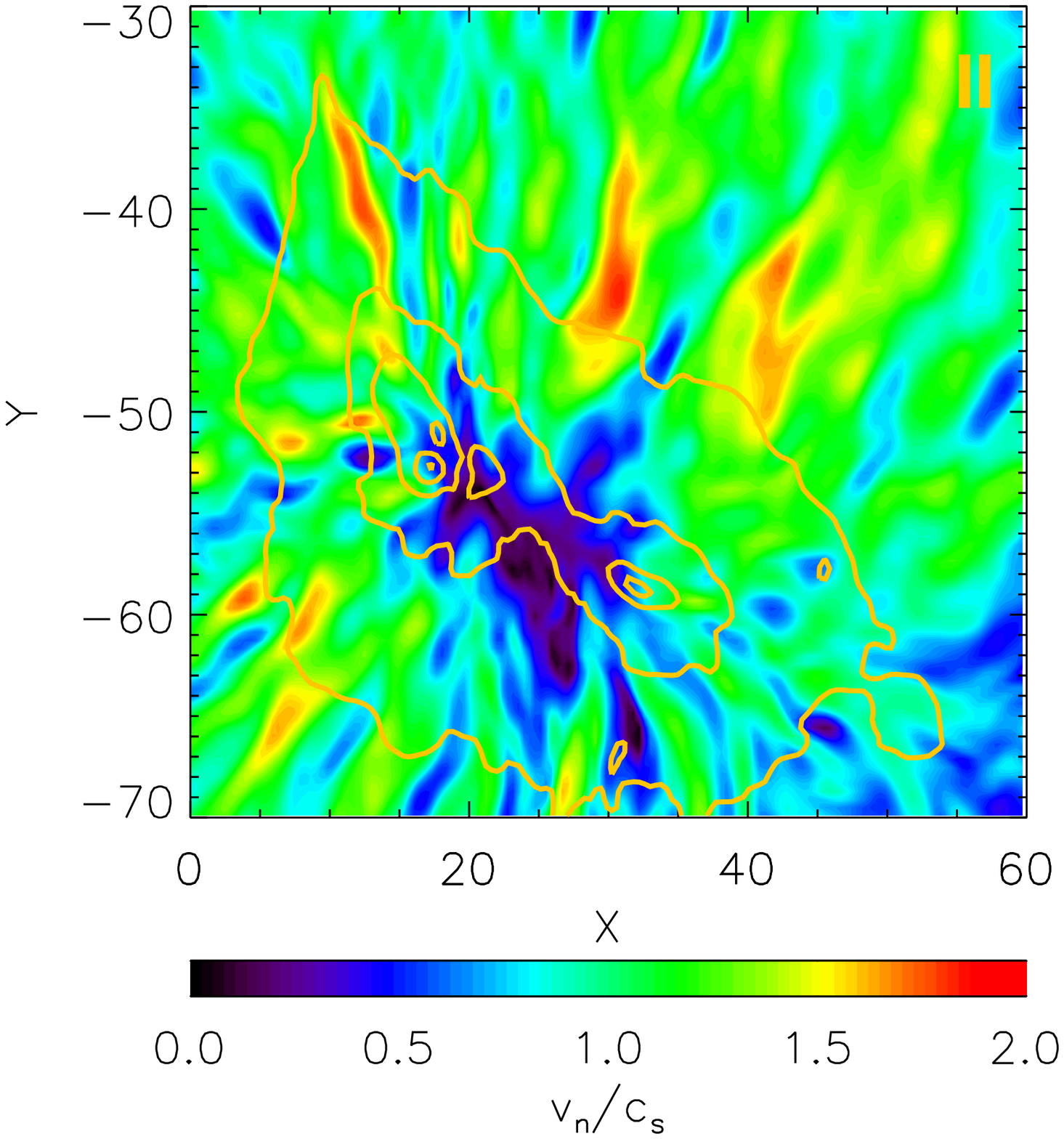}
\includegraphics[width = 0.33\textwidth]{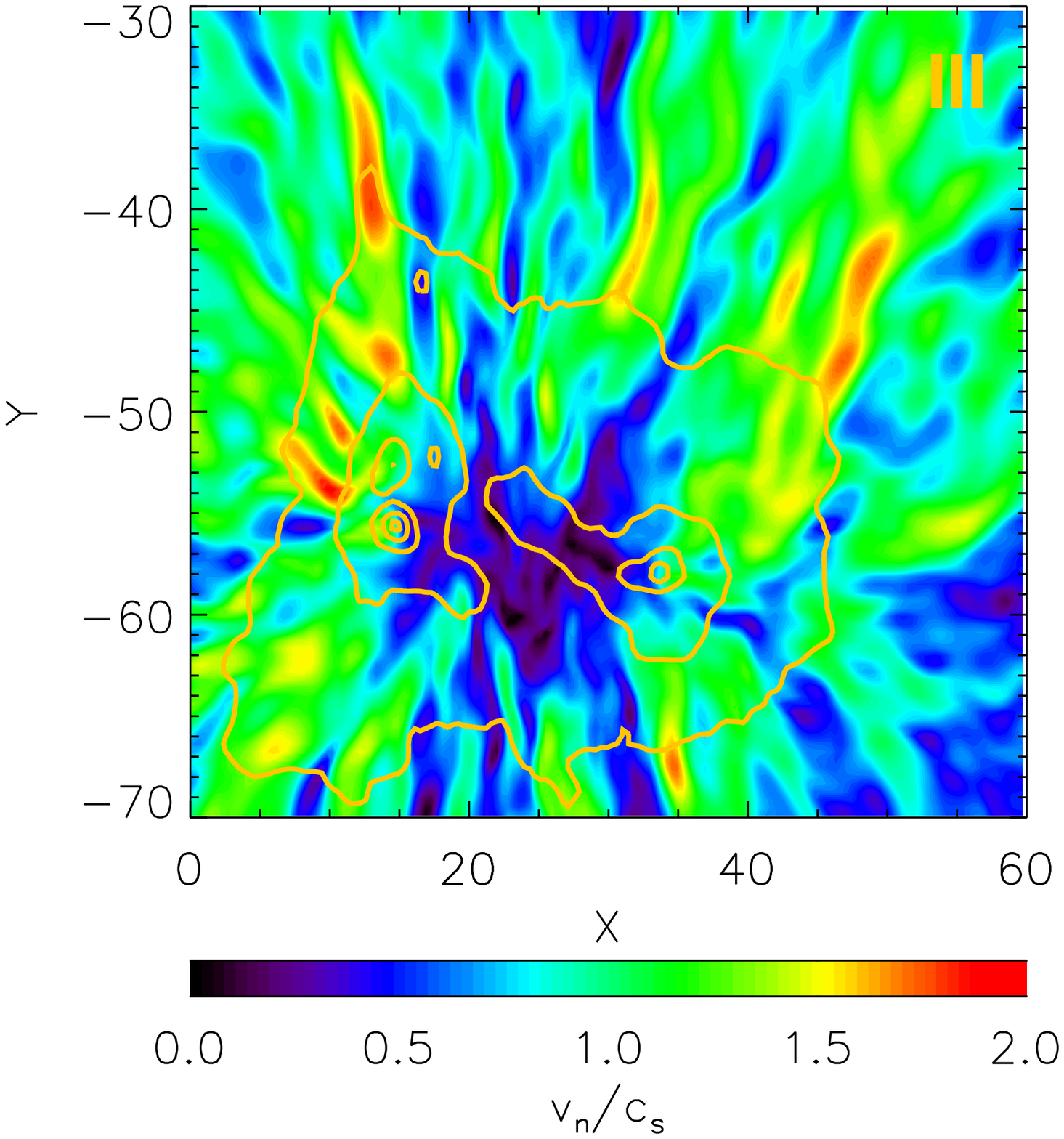}\\
\includegraphics[width = 0.33\textwidth]{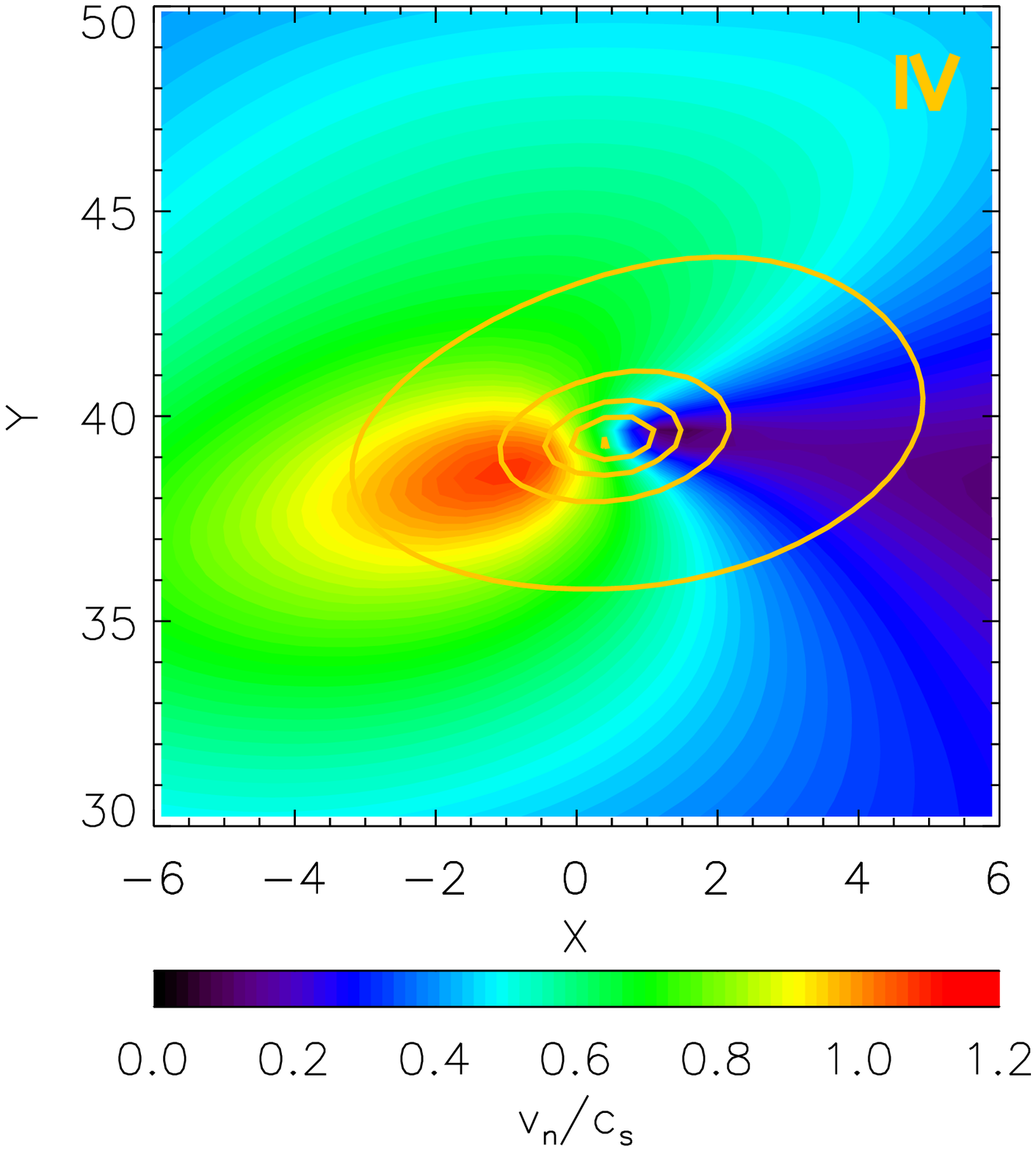}
\includegraphics[width = 0.33\textwidth]{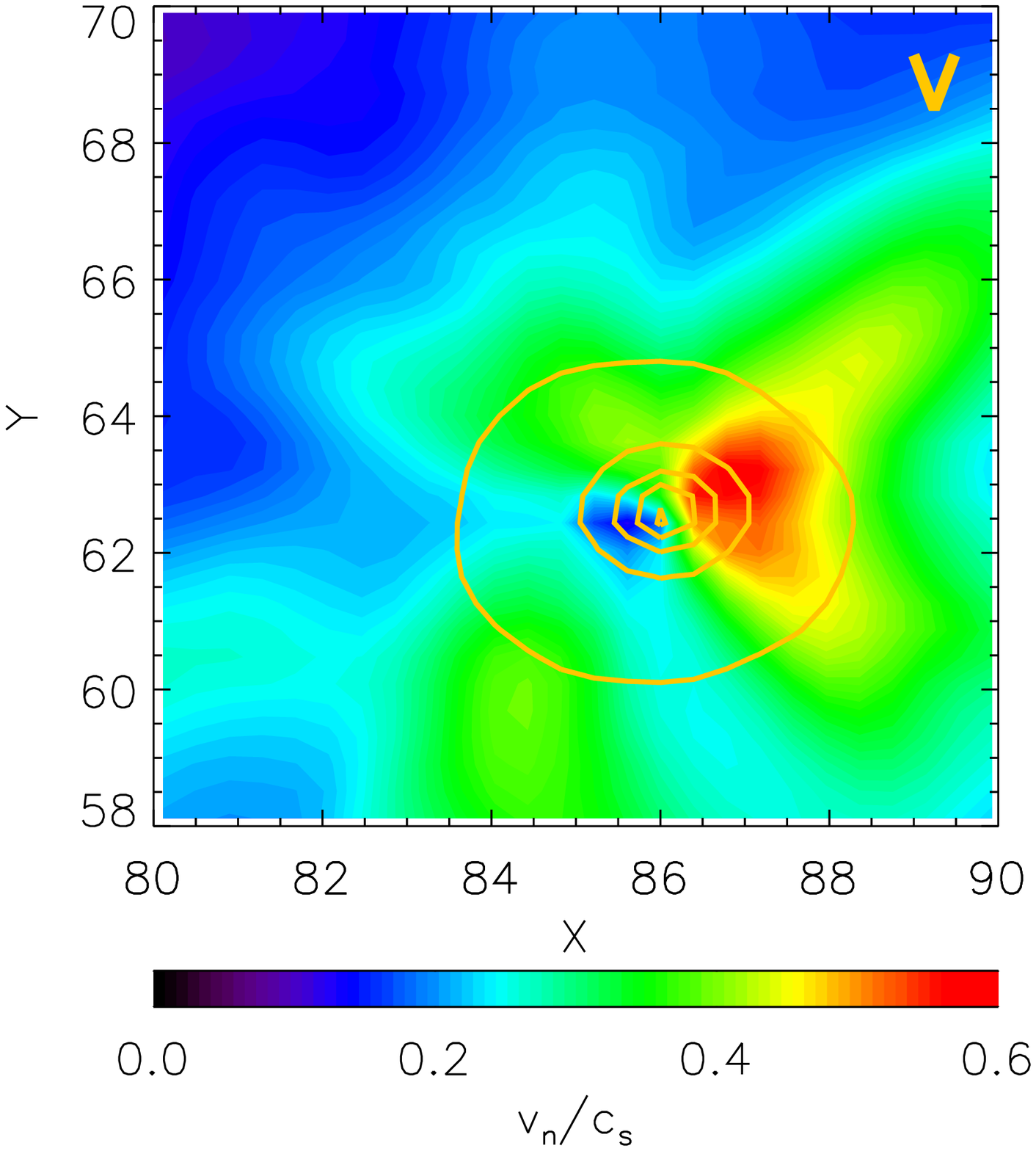}
\includegraphics[width = 0.33\textwidth]{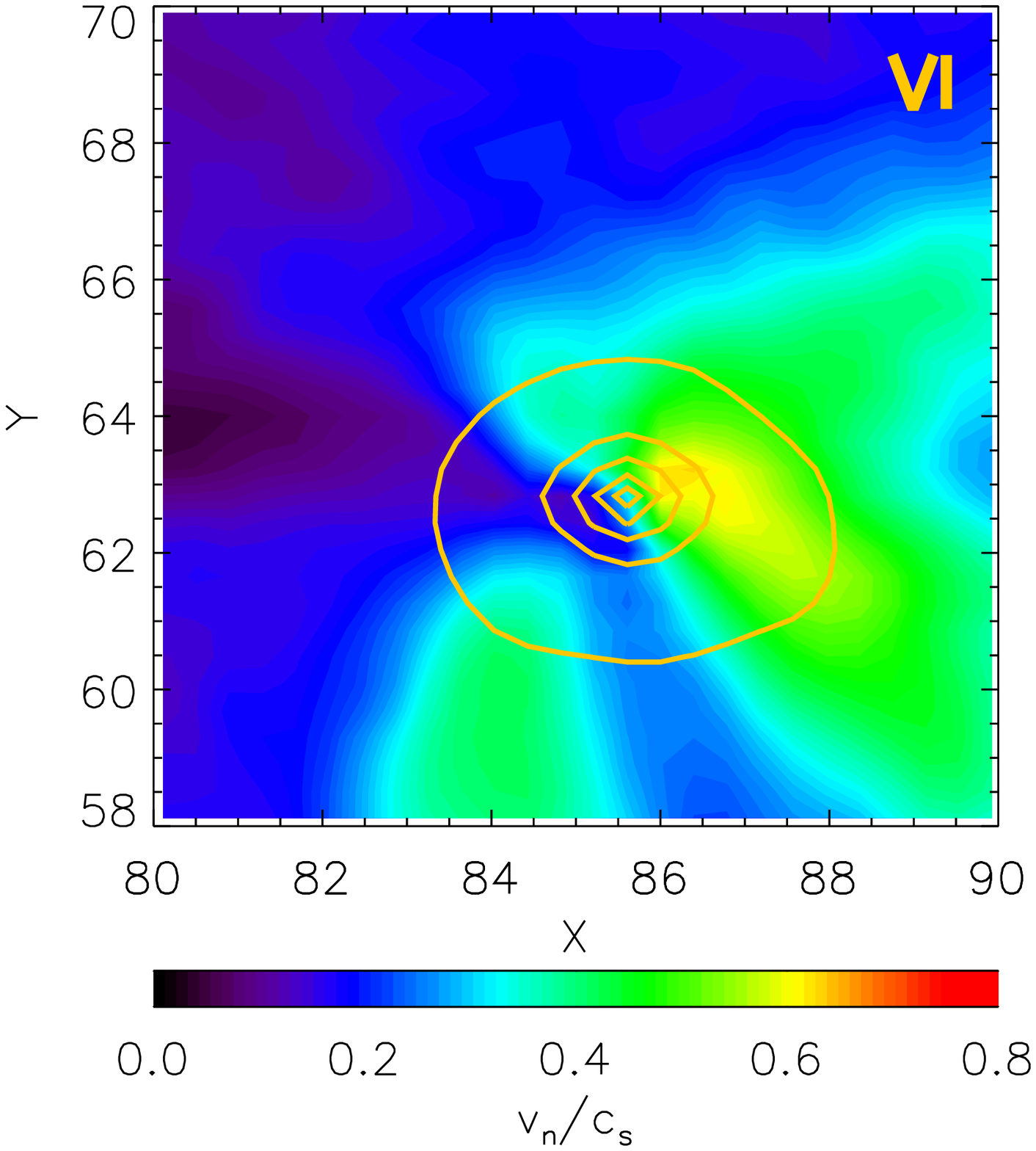}
\caption{Velocity maps at final time for each model with column density enhancement contours. Panels and contours depict the same models and levels as Figure~\ref{sigmaoverlays}, respectively.}
\label{velocity}
\end{figure*}

Observations of the velocity within a molecular cloud are restricted along the line of sight. Our simulations give us the opportunity to look at the velocity structure of the models within the plane of the sky. Figure~\ref{velocity} shows the velocity maps of the six models for the same regions depicted in Figure~\ref{sigmaoverlays}. The contours show the same column density enhancement levels as those depicted in Figure~\ref{sigmaoverlays} and are plotted to show the location of the clump/core structures in each model. Note the differences in the velocity ranges as denoted by the color bar for each panel. As shown, each model exhibits regions of high and low velocity, however we see that Models II and III have the largest velocity range while Models V and VI have the smallest. Looking at the models individually, we see that for Model I, the core regions exhibit the lowest velocity with two high velocity lobes on either side. For Models II and III, the addition of the ongoing perturbations changes the velocity structure dramatically. In these two cases, the low velocity region (hereafter referred to as the velocity valley) occurs in the center of the clump with high velocity streamers that exist on the outer edges. Specifically, the highest velocities tend to occur in the lower density gas. Finally, for the three models with the CR-only ionisation profile (Models IV - VI), the addition of perturbations still results in a chaotic velocity field, but to a lesser degree than in the step-like ionisation models. The larger degree of ambipolar diffusion throughout the simulations seems to result in a simpler velocity gradient with high velocities on one side of the core and low velocities on the opposite side. 
The distinct difference between the velocity structures formed under the different ionisation conditions is directly due to the assumed ionisation profile. Models that assume a step-like ionisation profile keep the low density regions at nearly flux frozen conditions, thus preventing collapse. Conversely, models that assume a CR-only ionisation profile allow for collapse to occur even at lower densities. Looking at Models II and V, for example, the difference in the velocity field between these two models is due to the fact that in Model II, the velocity magnitude increases in regions with steep gradients in the ionisation fraction. This velocity enhancement is caused by flows of material from high to low ionisation fraction regions as the ability for neutrals to slip past the magnetic field lines increases. With the lower ionisation fraction in Model V (or higher density regions of Model II), the gradients in the ionisation fraction are not as steep and therefore do not induce high velocities.

\begin{figure*}
\centering
\includegraphics[width = 0.33\textwidth]{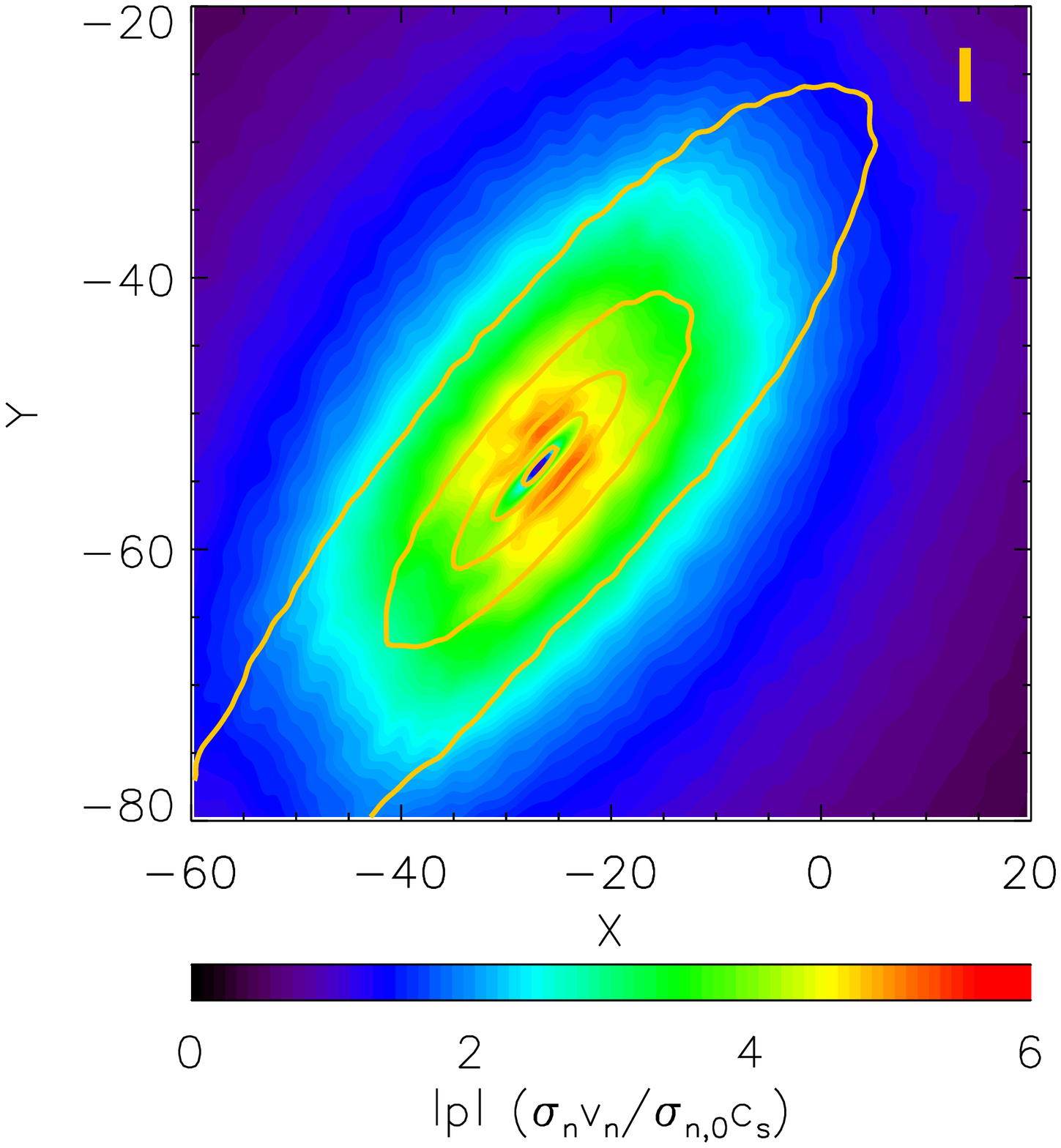}
\includegraphics[width = 0.33\textwidth]{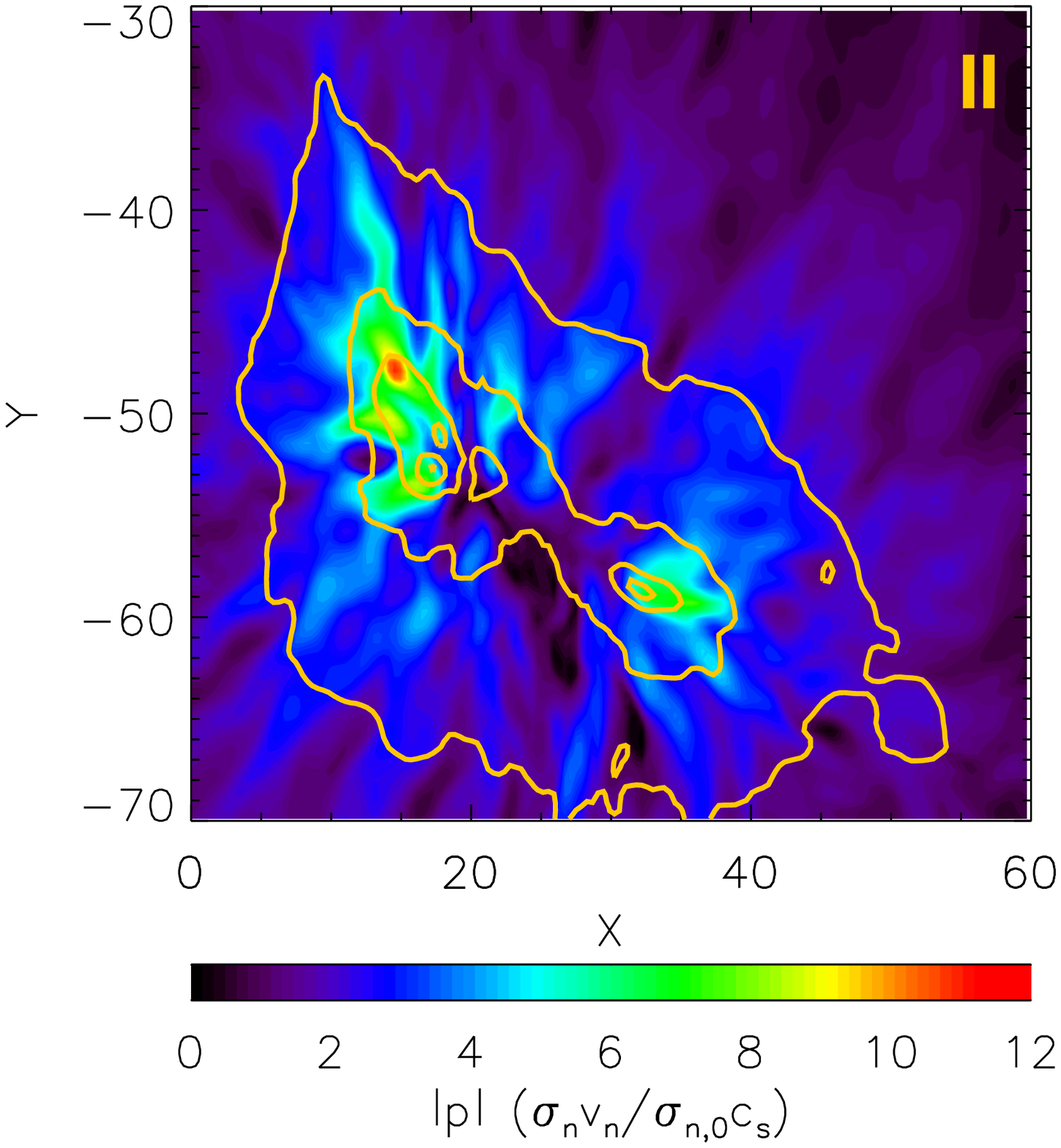}
\includegraphics[width = 0.33\textwidth]{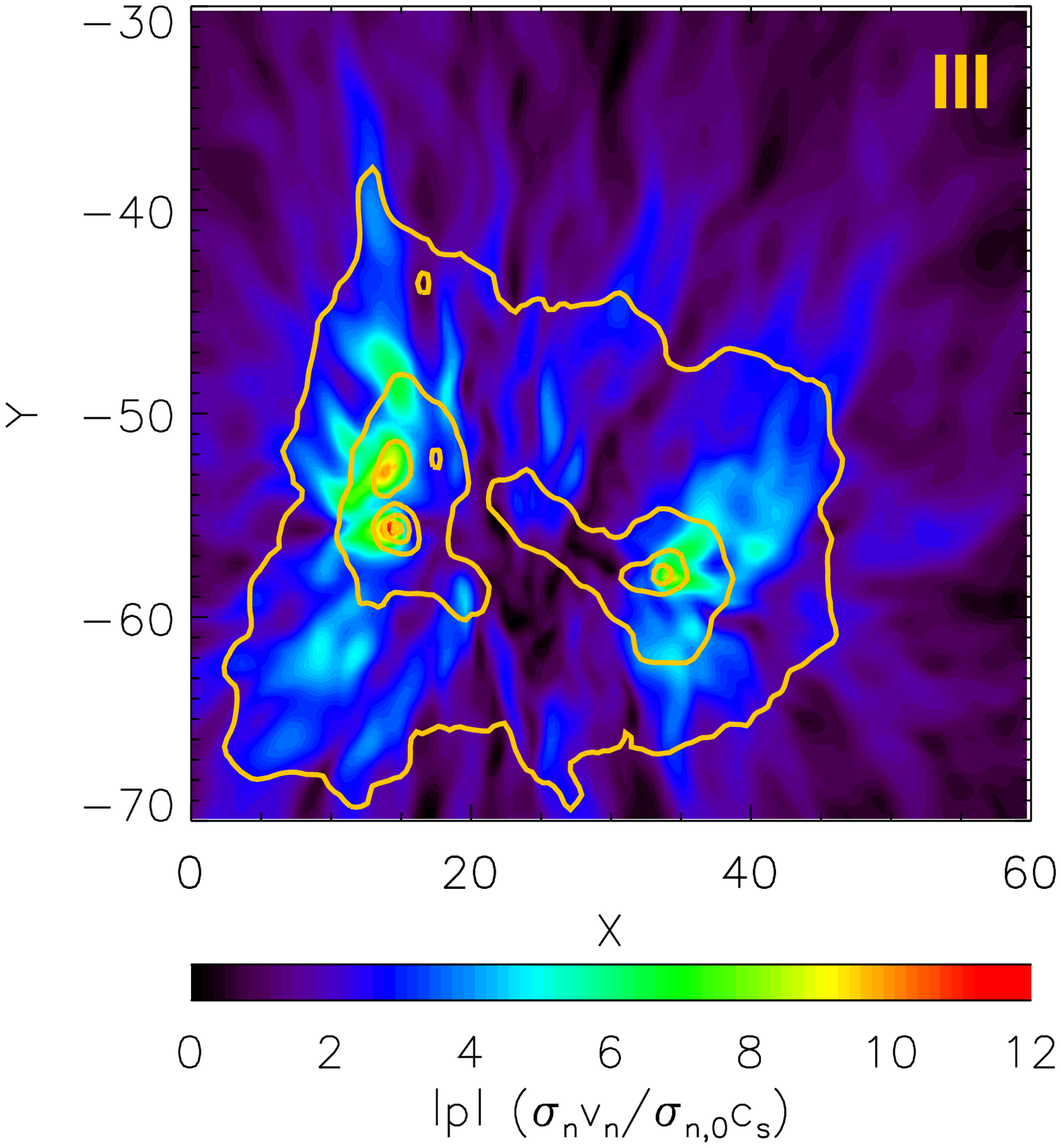}\\
\includegraphics[width = 0.33\textwidth]{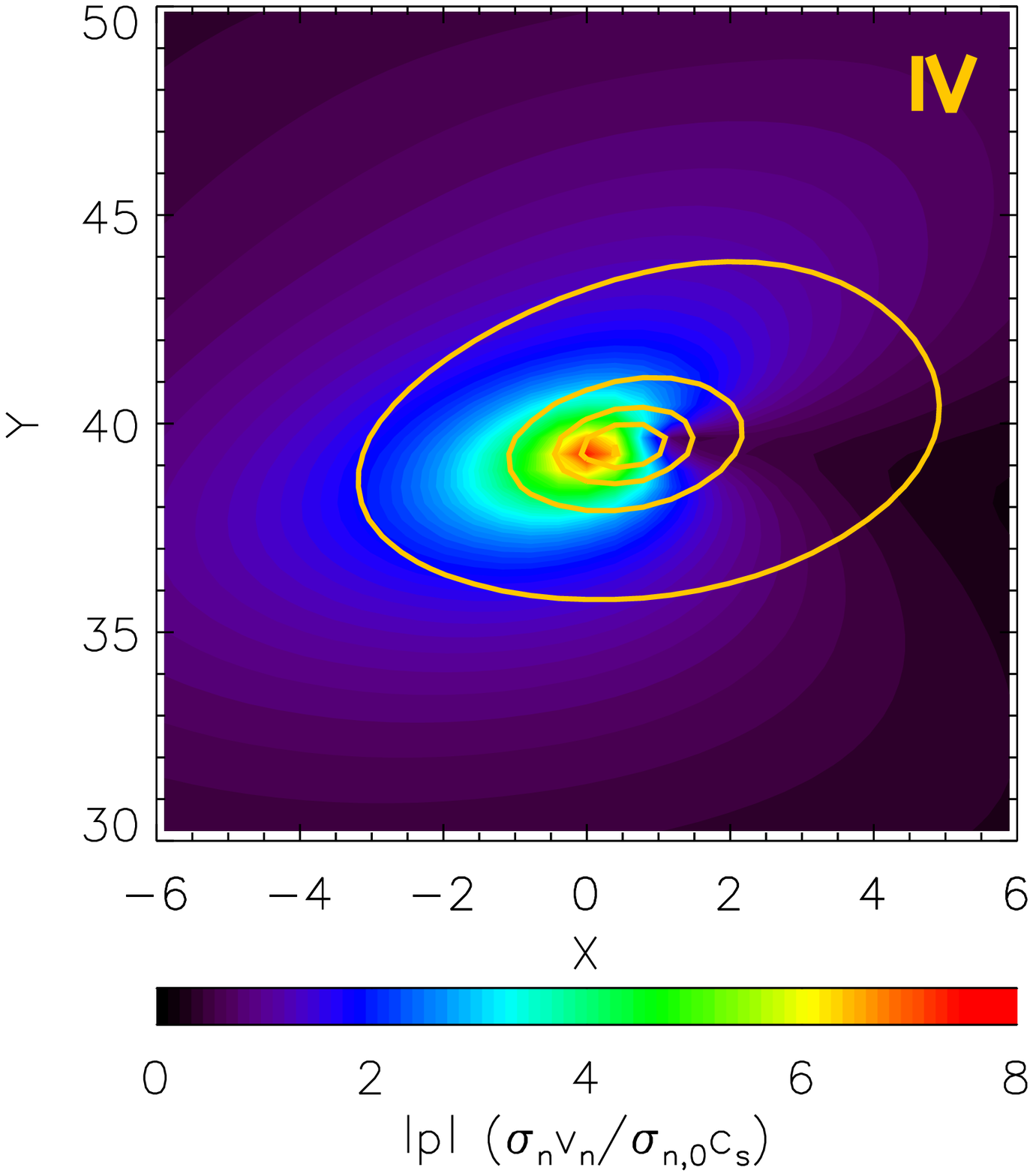}
\includegraphics[width = 0.33\textwidth]{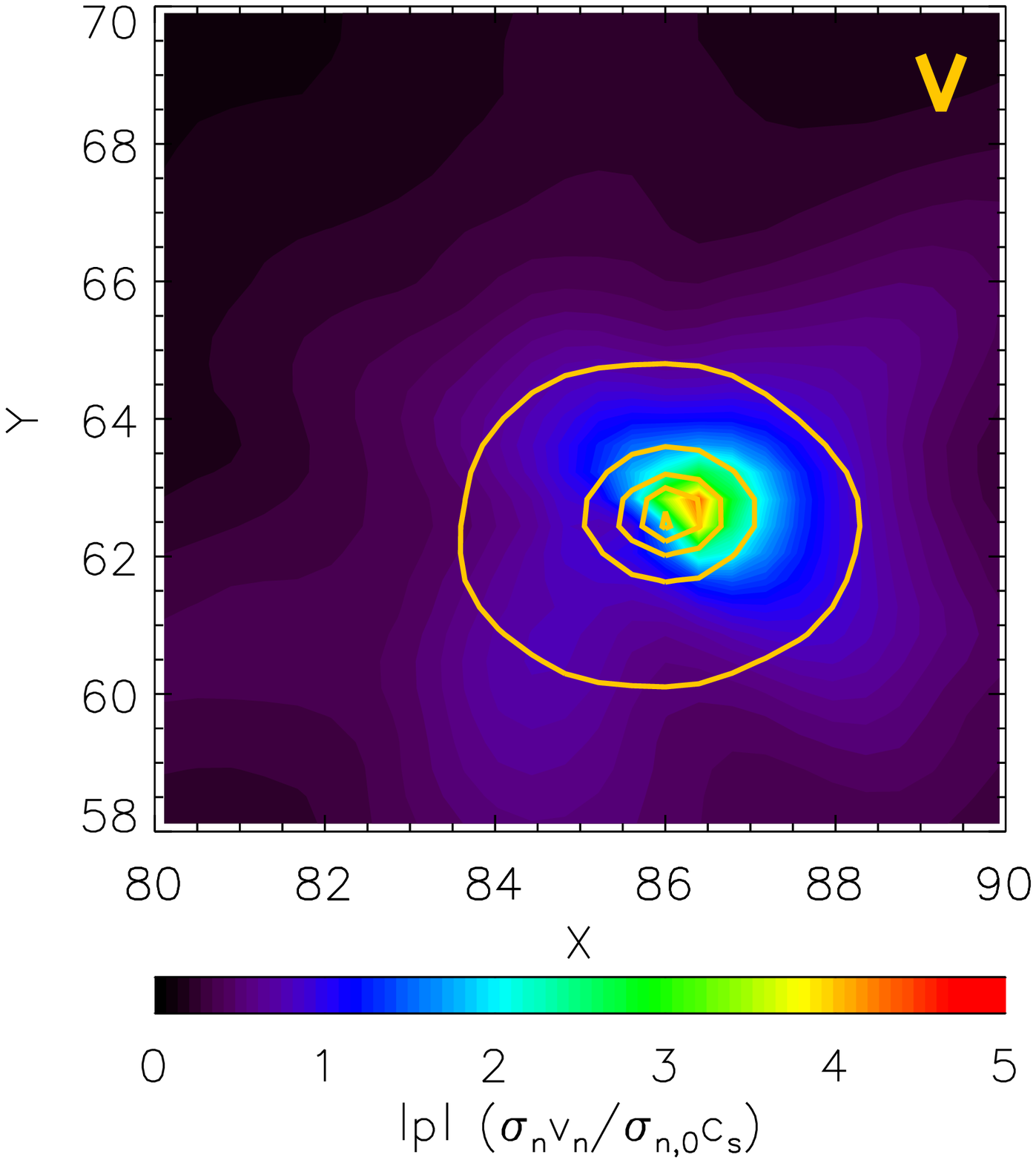}
\includegraphics[width = 0.33\textwidth]{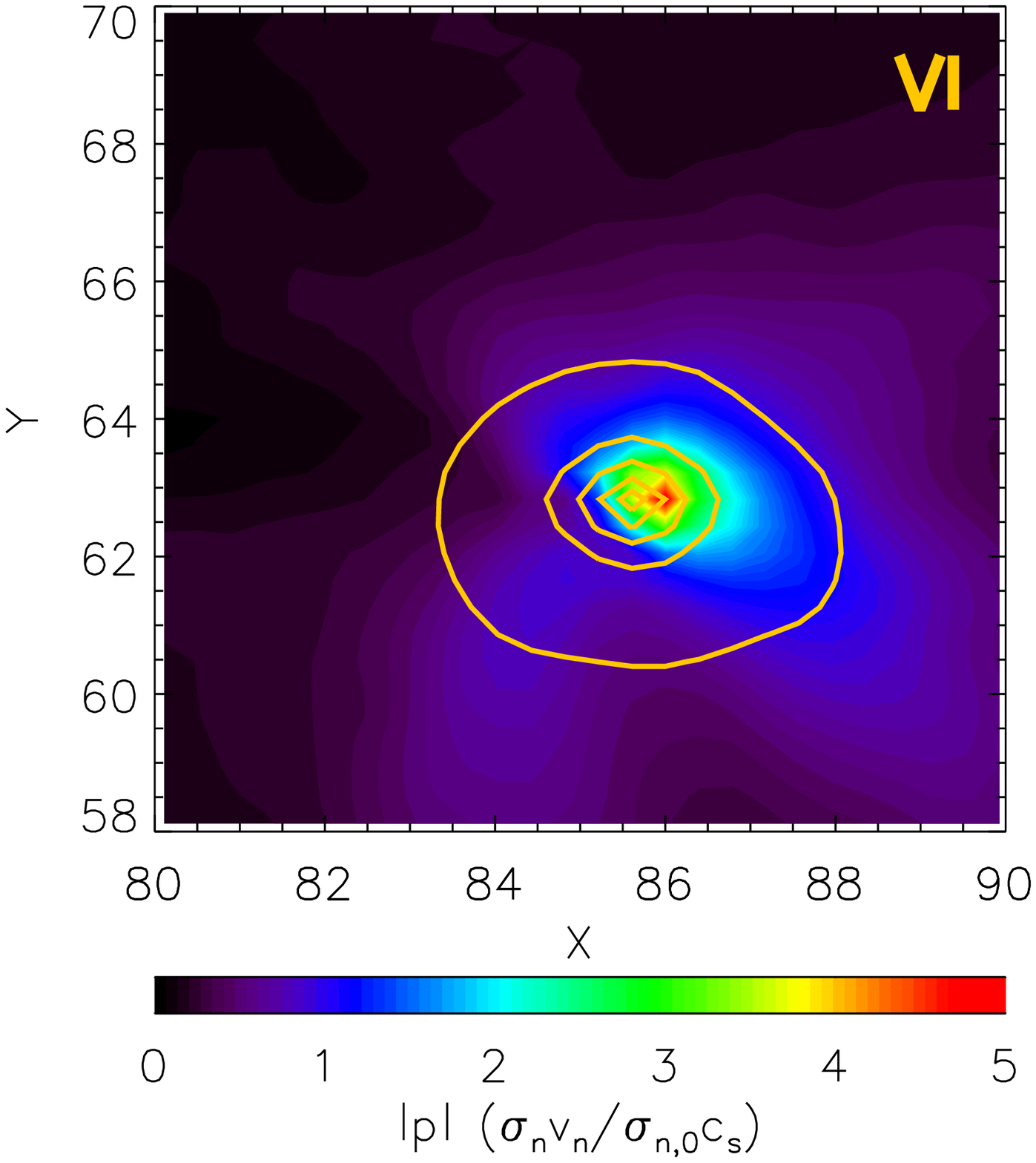}
\caption{Absolute value momentum maps ($|p|$) with column density enhancement contours. Panels and contours depict the same models and levels as Figure~\ref{sigmaoverlays}.}
\label{p_sigmaoverlays}
\end{figure*}

Figure~\ref{p_sigmaoverlays} shows the mass weighted velocity (momentum) maps with overlaid visual extinction contours for each of the six models. Again note the different color scales for each panel. Comparing the top row to the bottom, we see that the models with the step-like ionisation profile exhibit a larger momentum range than models with the CR-only ionisation profile. As with the velocity maps, the addition of on going perturbations acts to distort the momentum fields.  Looking closely at the panels for Models II and III (top middle and top right, respectively) we see that the largest momentum gradients seem to occur on the periphery of the cores.

\begin{figure*}
\centering
\includegraphics[width = 0.48\textwidth]{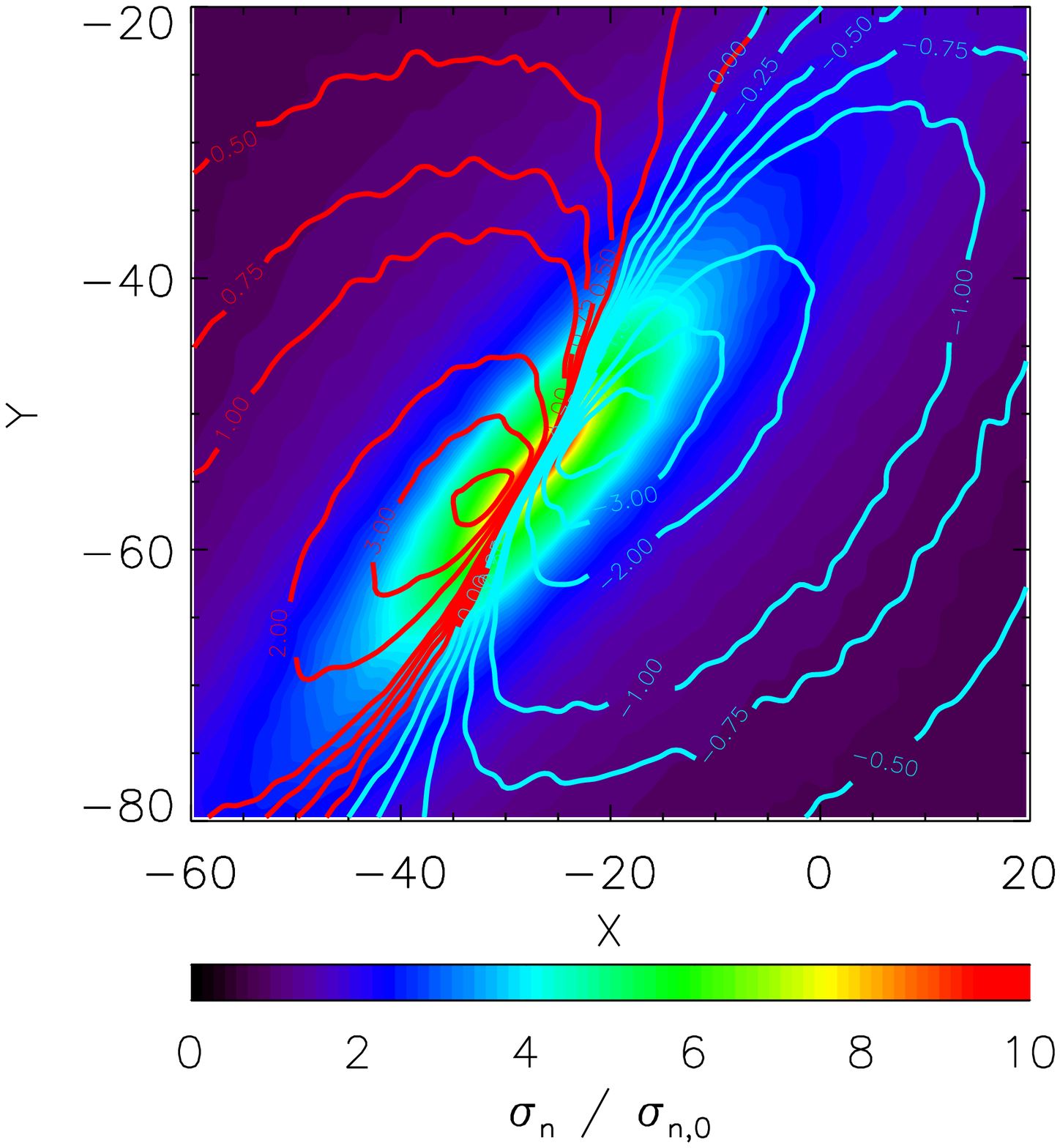}
\includegraphics[width = 0.48\textwidth]{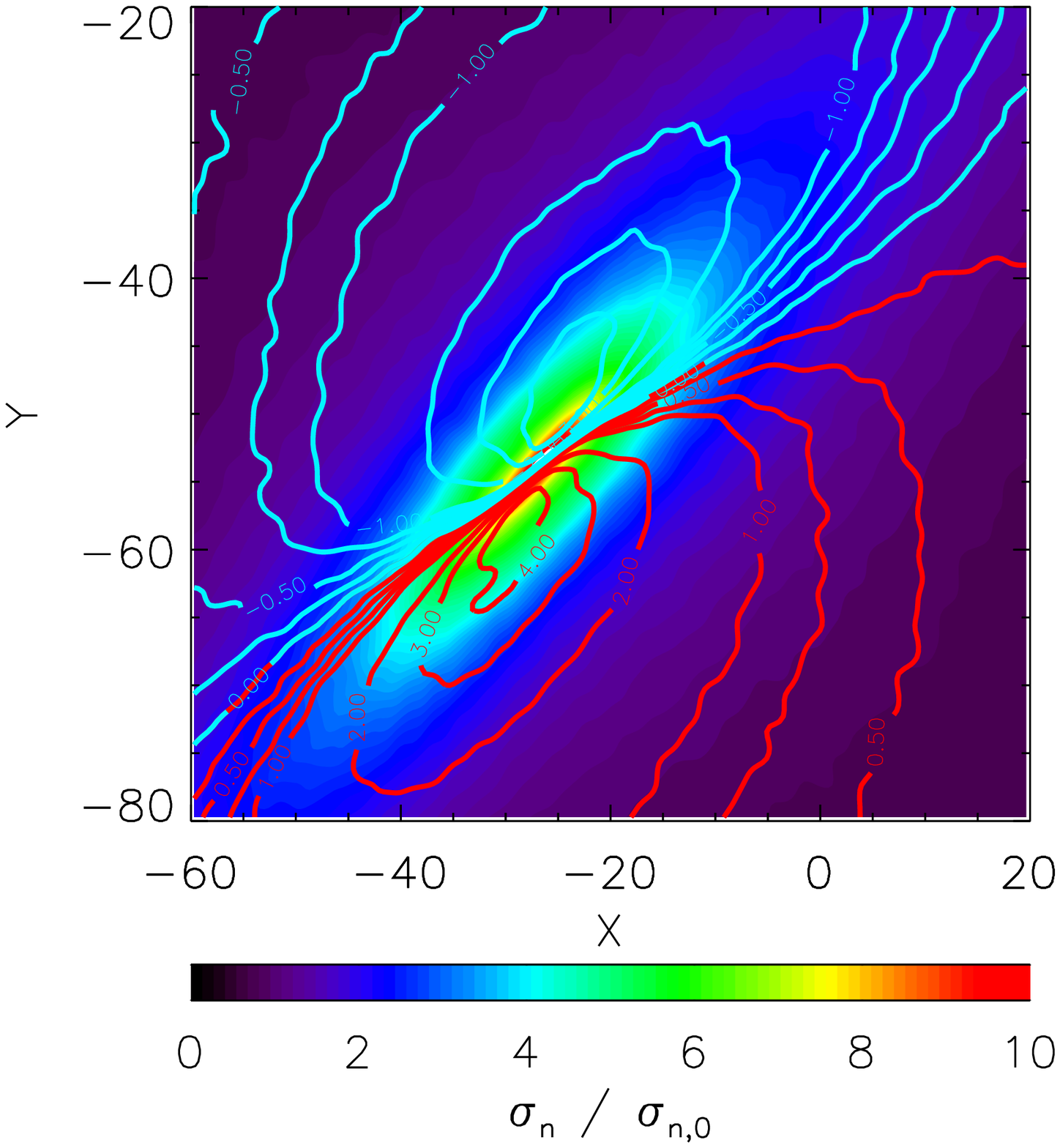}\\
\includegraphics[width = 0.48\textwidth]{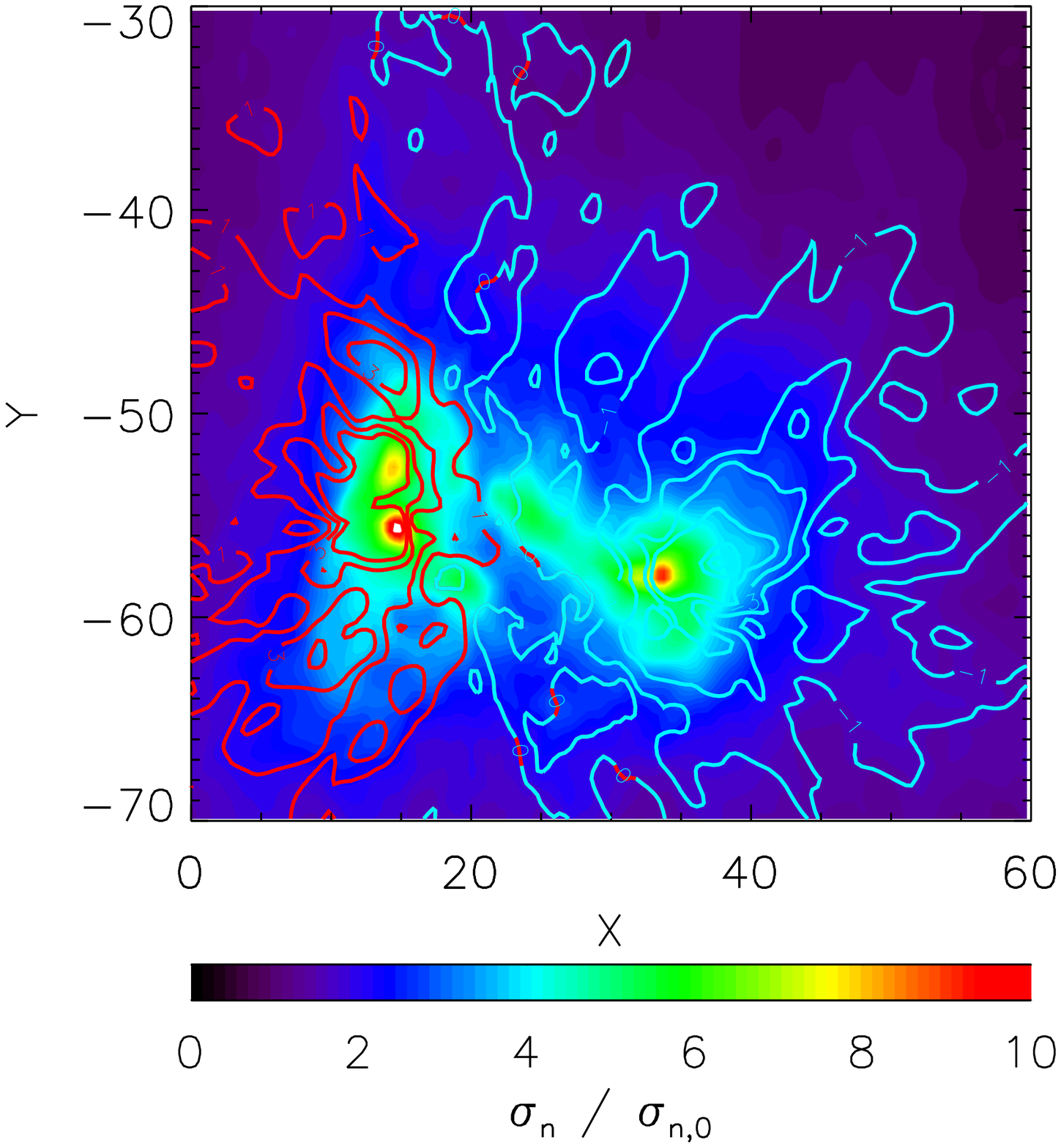}
\includegraphics[width = 0.48\textwidth]{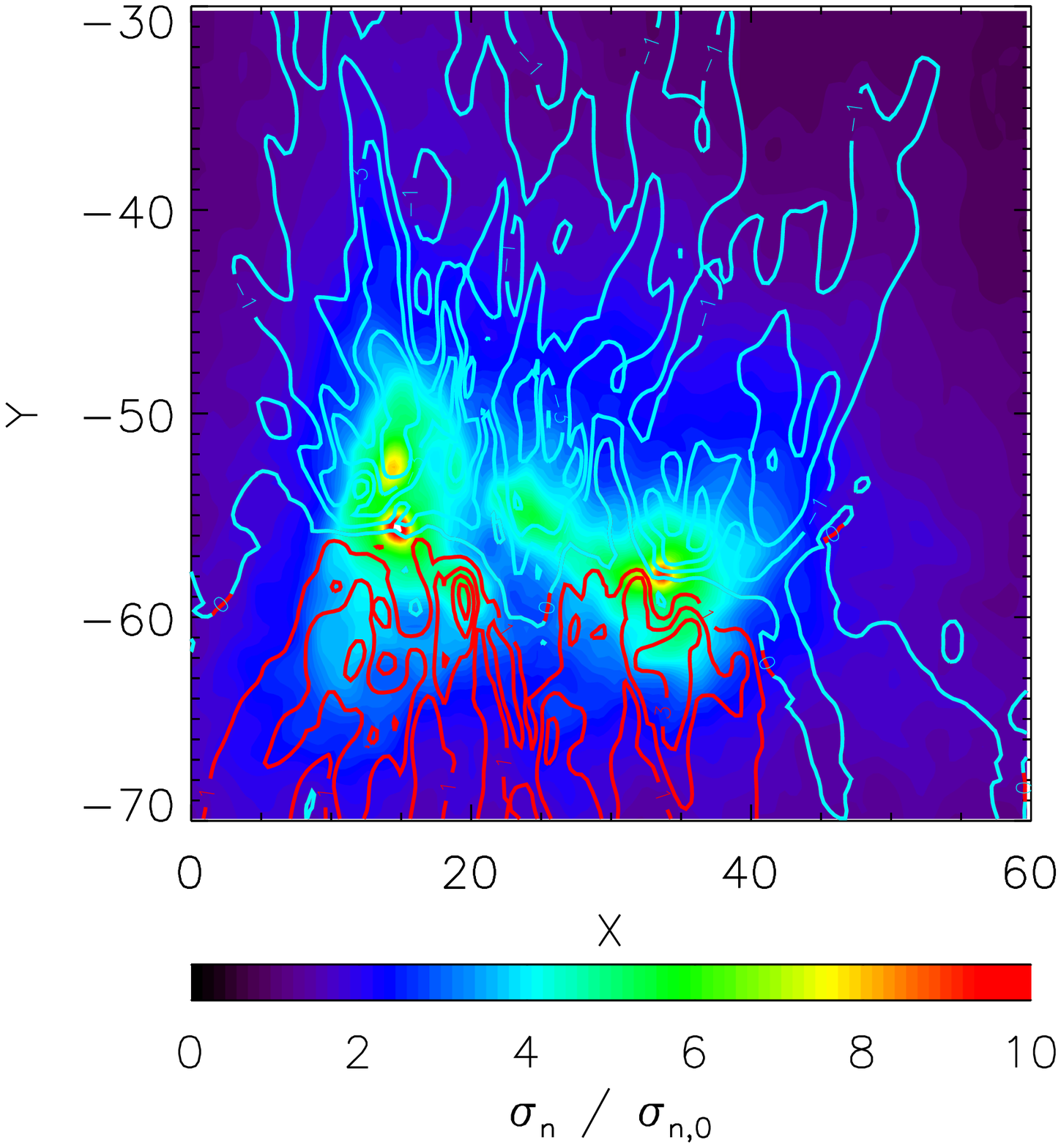}
\caption{Density enhancement maps with $p_{x}$ (left) and $p_{y}$ (right) momentum contours for Model I (top row) and Model III (bottom row). Cyan contours indicate motions in the negative direction (toward the observer)  and red contours indicate motions in the positive direction (away from observer). }
\label{sigma_poverlays}
\end{figure*}

Finally, Figure~\ref{sigma_poverlays} shows examples of density enhancement maps with overlaid $p_{x}$ and $p_{y}$ contours for Model I (top row) and Model III (bottom row). As depicted in all four panels, there exists an imaginary line where the contours switch from positive to negative. This indicates the convergence point for the gravitationally induced flows which form the clump/core. By comparing the two different momentum maps (Figures~\ref{p_sigmaoverlays}~\&~\ref{sigma_poverlays}) with the density enhancement and velocity maps in Figures~\ref{sigmaoverlays}~\&~\ref{velocity} respectively, we can determine the behaviour of the region. First, the components of the momenta show that the location of the clumps in all of the models occurs where the momentum switches signs from positive to negative. Second, the largest momentum gradients seem to occur on the periphery of the cores. On closer inspection, we see that these high momentum regions only occur on one side of the cores, indicating the direction of flow for the material from the surroundings onto the forming cores.

\section{Synthetic Spectra}

In order to see if the cores present in the simulations follow the three observed trends outlined in Section~\ref{aims}, we must perform spectral analysis on each of the models and as such must produce synthetic spectral observations of the cores and surrounding gas. Observations of star forming regions are three dimensional in nature, where the third dimension is the line of sight velocity ($V_{los}$). In principle, our simulations are two dimensional, however they do have a finite thickness in the $z-$ direction. Therefore, if we assume an observer is looking at the sheet edge-on, we can still extract synthetic spectra for two lines of sight, either along the $x-$ axis or along the $y-$ axis. For this rotation, we define the center of the full simulation region ($x,y = 0,0$) to be analogous to the center of a compass with $+y$ defined as north. Lines of sight that are parallel to the $y-$ axis originate from the southern extent of the simulation and terminate at the northern extent. Likewise, lines of sight parallel to the $x-$ axis originate from the western extent and terminate at the eastern extent. Based on this, we refer to these two orthogonal lines of sight as NS and EW, respectively, where the last letter indicates the location of the observer.

To compare the kinematic properties of the simulated cores to the observed properties outlined in Section~\ref{aims}, we need synthetic spectra of the neutral and ionic gas components of both the cores and surrounding low density gas. The inclusion of ambipolar diffusion in the simulations allows us to examine the synthetic spectra for both the neutral and ionic components of the medium. When comparing to observations, we must choose the appropriate tracers that our simulations will correspond to. Based on the termination conditions of the simulations, the final range of visual extinctions within our models runs from $<$~1 mag to just above 10 mag, which corresponds to column densities $N = 9\times 10^{20}$~cm$^{-2}~-~9\times 10^{21}$~cm$^{-2}$. As we are restricting our analysis to the clump-core regions, we must assume tracers that are appropriate for the densities within these regions. Ammonia (NH$_{3}$) and Diazenylium (N$_{2}$H$^{+}$) are excellent neutral and ionic tracers respectively for regions with molecular hydrogen densities in the range of 10$^{4}$ cm$^{-3}$ - 10$^{5}$ cm$^{-3}$ and $A_{V} = 3 - 9.5$ mag \citep{Tafalla2002, Caselli2002a}. As shown by the panels in Figure~\ref{sigmaoverlays}, the clump/core regions are encompassed within the $A_{V} = 2$ mag contour with the most structure appearing within the $A_{V} = 4$ mag contour where the average volume density is above $4.4\times 10^{3} \rm~cm^{-3}$, thus indicating that the majority of our region of interest would indeed be detected by these two tracers. For the low-density gas, we assume the majority of the neutral particles are carbon monoxide (CO) and the majority of ions are H$^{13}$CO$^{+}$. No chemistry is included in the models, so constant abundances of the various species are assumed, as well as optically thin conditions. The effects of the inclusion of a simplified chemistry and radiative transfer will be discussed in a future paper. 

\subsection{Linewidth and Centroid Velocity Analysis}
\subsubsection{Method}

With the above points in mind, we construct synthetic spectra by assuming Gaussian line shapes for each pixel such that the line of sight velocity in each pixel (i.e., $V_{x}$ or $V_{y}$ depending on orientation) is the mean velocity for the pixel. The width of the line is dependent on the thermal and non-thermal velocities, i.e.,
\begin{equation}
FWHM = \sqrt{\Delta V_{NT}^{2} + \Delta V_{T}^{2}},
\end{equation}
 where $\Delta V_{T}^{2} = 8 \ln 2~c_{s}^{2}$ is the thermal velocity component and $\Delta V_{NT}^{2} = \Delta V_{obs}^{2} - \Delta V_{T}^{2}$ is the non thermal velocity component with $\Delta V_{obs}$ corresponding to the observed velocity \citep{Myers1991}. For each pixel, we assume that they are small enough such that they are thermalised, i.e., there is no non-thermal component, so that observed non-thermal motions will reflect the kinematics. Based on this assumption, the width of the Gaussian is 
\begin{equation}
FWHM = \sqrt{\Delta V_{T}^{2}} = \sqrt{8 \ln 2~c_{s}^{2}}.
\end{equation}  
The height of the Gaussian is given by the FWHM scaled by the neutral or ionic column density. The total spectral line is then constructed by summing up all individual components along the line of sight.

For our analysis, we look at spectra for lines of sight both on and off source. For the on-source spectra, we create spectra for the two perpendicular lines of sight (NS and EW) discussed in the previous section. For each line of sight, we create four separate spectra: a low density component and core component for both the neutral and ionic gas, respectively, assuming that the high and low density components are traced by the four molecules discussed above. For this analysis, the low density gas (LDG) is defined as regions along the los with visual extinction $2~<~A_{V}~<~7$ mag and the cores are defined as regions along the los with visual extinction $A_{V} > 7$ mag. The sources for which we create these spectra are listed in Table~\ref{coordinates}. For each model, the MC designation refers to the ``Main Core'' which is the most evolved core (i.e., the one that caused the simulation to terminate; see Section~\ref{model}). Models II and III also exhibit a second well-evolved core in the south east (SE) region of the clump. For these models, we investigate this core in addition to the MC in order to determine how the spectra may change with contamination from gas originating from other regions of the clump.

\begin{deluxetable}{lcrr}
\tablecaption{Core Designations}
\tablewidth{0pt}
\tablehead{
\colhead{Model} & \colhead{Designation}  & \colhead{$x$} & \colhead{$y$}
}
\startdata
I   & MC & -26.5 & -53.5 \\    
II  & MC &  17.6 & -52.0 \\
II  & SE &  32.0 & -57.8 \\
III & MC &  15.2 & -55.1 \\ 
III & SE &  34.0 & -57.4 \\
IV  & MC &   0.4 &  39.1 \\
V   & MC &  85.5 &  62.1 \\
VI  & MC &  85.2 &  62.5 
\enddata
\label{coordinates}
\end{deluxetable}

\subsubsection{Results: Core and LDG Properties}

\begin{figure*}
\includegraphics[width = 0.76\textwidth, angle=-90]{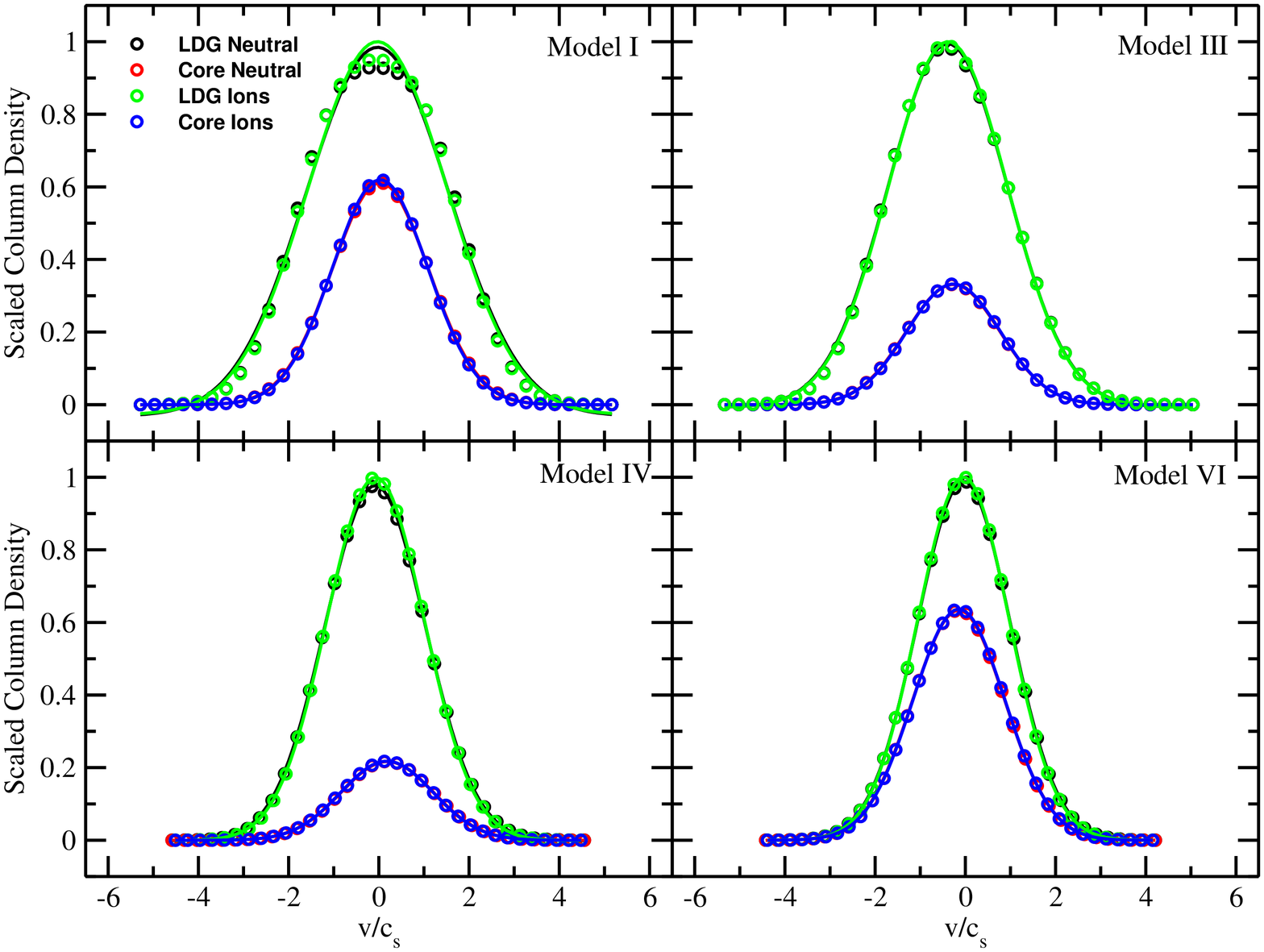}
\caption{Examples of synthetic spectra for Model I (upper left), Model III (upper right), Model IV (lower left), and Model VI (lower right). Plots show the synthetic lines and Gaussian fits for the low density gas (outer Gaussian) and core (inner Gaussian) as indicated by the legend. Symbols indicate the ``observed'' values while the solid lines show the Gaussian fits. All spectra are scaled such that the strongest line in each panel corresponds to a scaled column density of unity. All panels show a NS los through the main core in each individual model (see Table~\ref{coordinates}) and assume a low density gas range 2 mag $< A_{V} <$ 7 mag and a lower core threshold of $A_{V} = 7$ mag.}
\label{spectra}
\end{figure*}

Figure~\ref{spectra} shows examples of the synthetic spectra for Models I (upper left), III (upper right), IV (lower left), and VI (lower right). Each panel shows the spectra for the low-density gas (outer Gaussian) and the component of the core (inner Gaussian) for the assumed visual extinction ranges. The open circles show the synthetically generated spectra for the neutral low density gas (black), neutral core gas (red), ionic low density gas (green) and ionic core gas (blue). The correspondingly colored solid lines show the Gaussian fits for each of the four species. All spectra were scaled such that the strongest line in each panel corresponds to a column density of unity. The data represented by the black and red symbols (low density and core neutral gas, respectively) are not visible on the plot given that they are obscured by the green and blue symbols (low density and core ionic gas, respectively). This indicates that for both ionisation profiles, the motions of the ions and neutrals are very similar. Comparing the models that assume a step-like ionisation profile (top row) to those with a cosmic ray only profile (bottom row) we see that both the core and low density envelope within models with the step-like ionisation profile exhibit large non-thermal contributions to the Linewidth while the CR only models only show evidence of slight non-thermal contributions.

\begin{figure*}
\includegraphics[width = 0.76\textwidth, angle = -90]{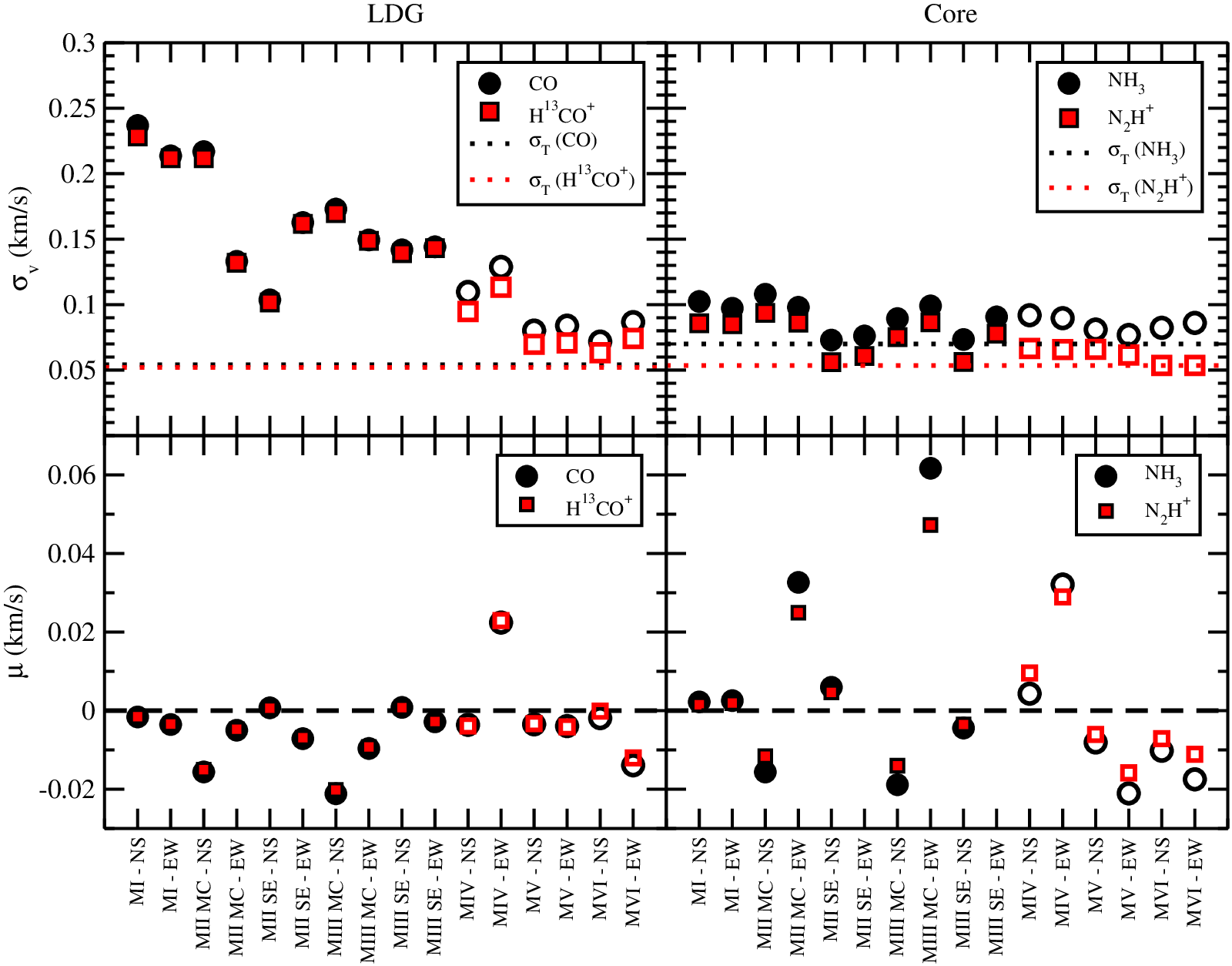}
\caption{Resulting Gaussian parameters for both neutral particles and ions for all models and lines of sight. Symbols depict values for the assumed low density gas (LDG) or core tracers as indicated by the legends. The solid symbols depict models with the step-like ionisation profile while open symbols depict models with the CR-only ionisation profile. The LDG includes gas within the visual extinction range 2 mag $<~A_{V}~<$ 7 mag, while core gas includes gas with $A_{V}~>~7$ mag. Top panels: velocity dispersion ($\sigma_{V}$). Bottom panels: centroid velocity ($\mu$). The black and red dotted lines indicate the thermal variance ($\sigma_{T}$) for each respective molecule assuming a temperature of 10 K. The dashed line shows $\mu = 0$ for visual reference.  }
\label{gaussianparams}
\end{figure*}

Figure~\ref{gaussianparams} shows the results of fitting Gaussians to each of the synthetic spectra. Panels show the velocity dispersion ($\sigma_{v}$, top) and the centroid velocity ($\mu$, bottom) for the LDG (2 mag $<~A_{V}~<$ 7 mag, left) and core gas ($A_{V}~>~7$ mag, right). Here we have scaled the fitted parameters to the dimensional values assuming the mean mass for the non-thermal component and the mass of the molecular tracer for the thermal component. As discussed above, we assume the low density neutral gas, low density ionic gas, neutral core gas and ionic core gas are traced by CO, H$^{13}$CO$^{+}$, NH$_{3}$ and N$_{2}$H$^{+}$, respectively. The solid symbols depict models with the step-like ionisation profile while lighter symbols depict models with the CR-only ionisation profile. The black and red dotted lines in the upper panels indicate the thermal velocity dispersion for each molecule assuming a temperature of 10 K. The dashed line in the lower panels depicts the $\mu = 0$ line for visual reference. Focusing on the top panels first, we note that in general, the low-density gas has a larger dispersion than the core gas. This is especially evident for the models with the step-like ionisation profile. This is likely due to the fact that with the step-like ionisation profile, the low-density gas is kept near flux freezing throughout the entire simulation. As such, the non-thermal component added by the microturbulent perturbations is not able to dissipate as efficiently as in the CR-only models. The neutral gas is expected to have slightly larger dispersions than the ionic component (see  next paragraph). This should be reflected in larger $NH_{3}$ line widths compared to the $N_{2}H^{+}$ line widths, if the two species do indeed trace the same gas \citep[see][for possible exceptions]{Friesen2010, Tafalla2004}.  Looking at the bottom panels, we see no discernible trend between the LDG and core gas. The centroid velocity of ions and neutrals within the high-density gas can differ significantly depending on the line of sight and model.

\begin{figure}
\includegraphics[width = 0.38\textwidth, angle = -90]{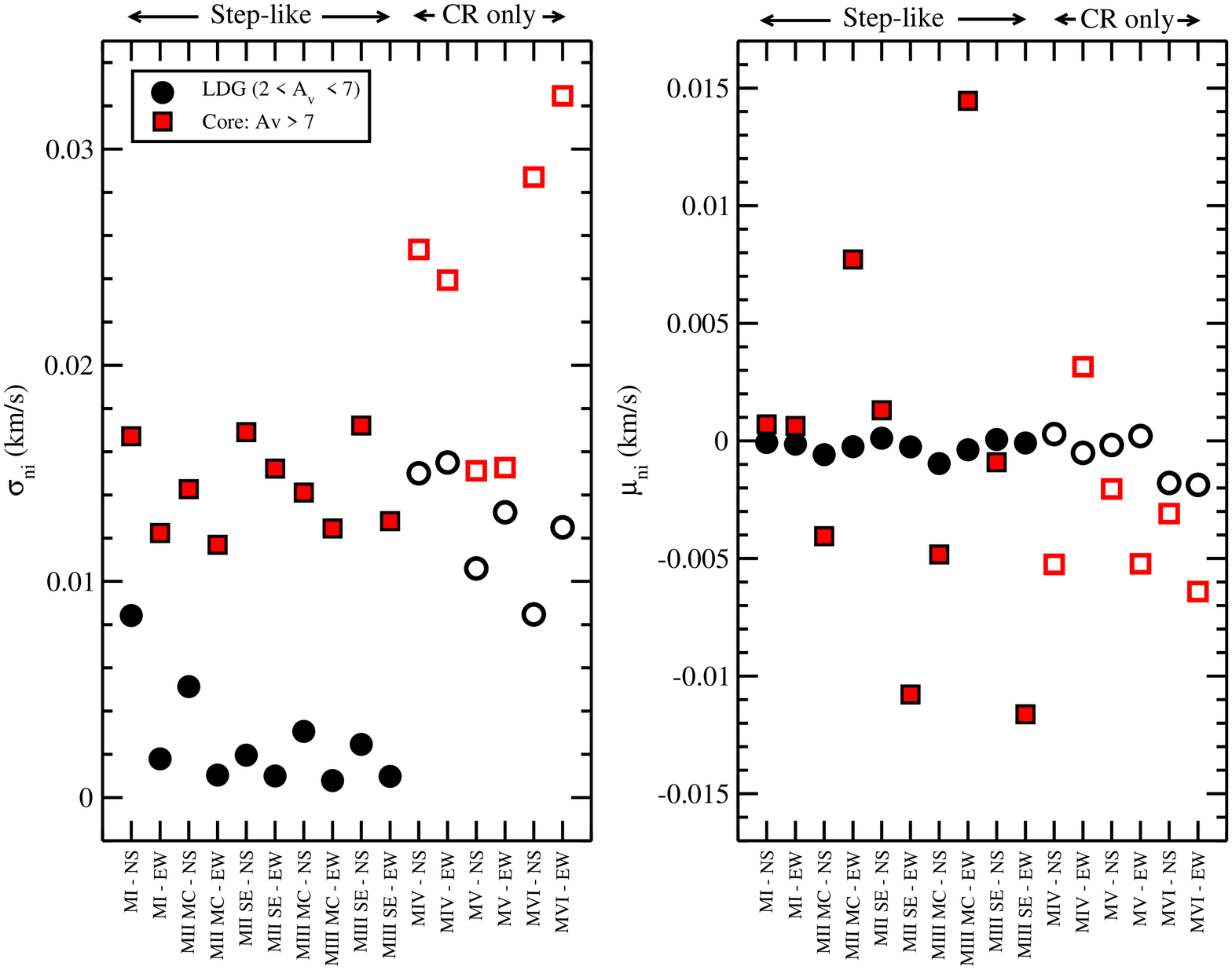}
\caption{Difference between neutral and ions for low density gas and cores. Symbols show the difference (neutral - ion) for the visual extinction ranges indicated in the legend. Solid symbols indicate models with the step-like ionisation profile while open symbols indicate models with CR-only ionisation profile. Left Panel: Difference in the velocity dispersion. Right Panel: Difference in the centroid velocity.}
\label{comparison1}
\end{figure}

Figure~\ref{comparison1} highlights the difference between the Gaussian parameters of the neutrals and ions for each of the core lines of sight studied. The left panel shows the difference between the velocity dispersion of the neutrals and ions ($\sigma_{ni}$) while the right panel shows the difference between the centroid velocity of the neutrals and the ions ($\mu_{ni}$). The solid symbols depict models with the step-like ionisation profile while the lighter symbols depict models with the CR-only ionisation profile. Looking at the left panel first, we see that for all models, the difference between the neutrals and ions is always positive. This indicates that the neutrals have larger velocity dispersions than the ions within both the LDG and cores regardless of the ionisation profile. This is because gravitationally driven motions affect neutrals more strongly than the ions. In addition, we see that in all cases, the difference between the neutrals and ions is larger for the cores than for the low-density gas. Also evident is a distinct split between the two different ionisation profiles. For both ionisation profiles, there does not seem to be a discernible trend with increasing frequency of perturbations. Looking at the right hand panel, we see that the difference between centroid velocities of the neutral particles and ions for the LDG is small while the difference for the core gas shows larger deviations relative to the low density gas for some lines of sight.

\begin{figure}
\centering
\includegraphics[width = 0.37\textwidth,angle=-90]{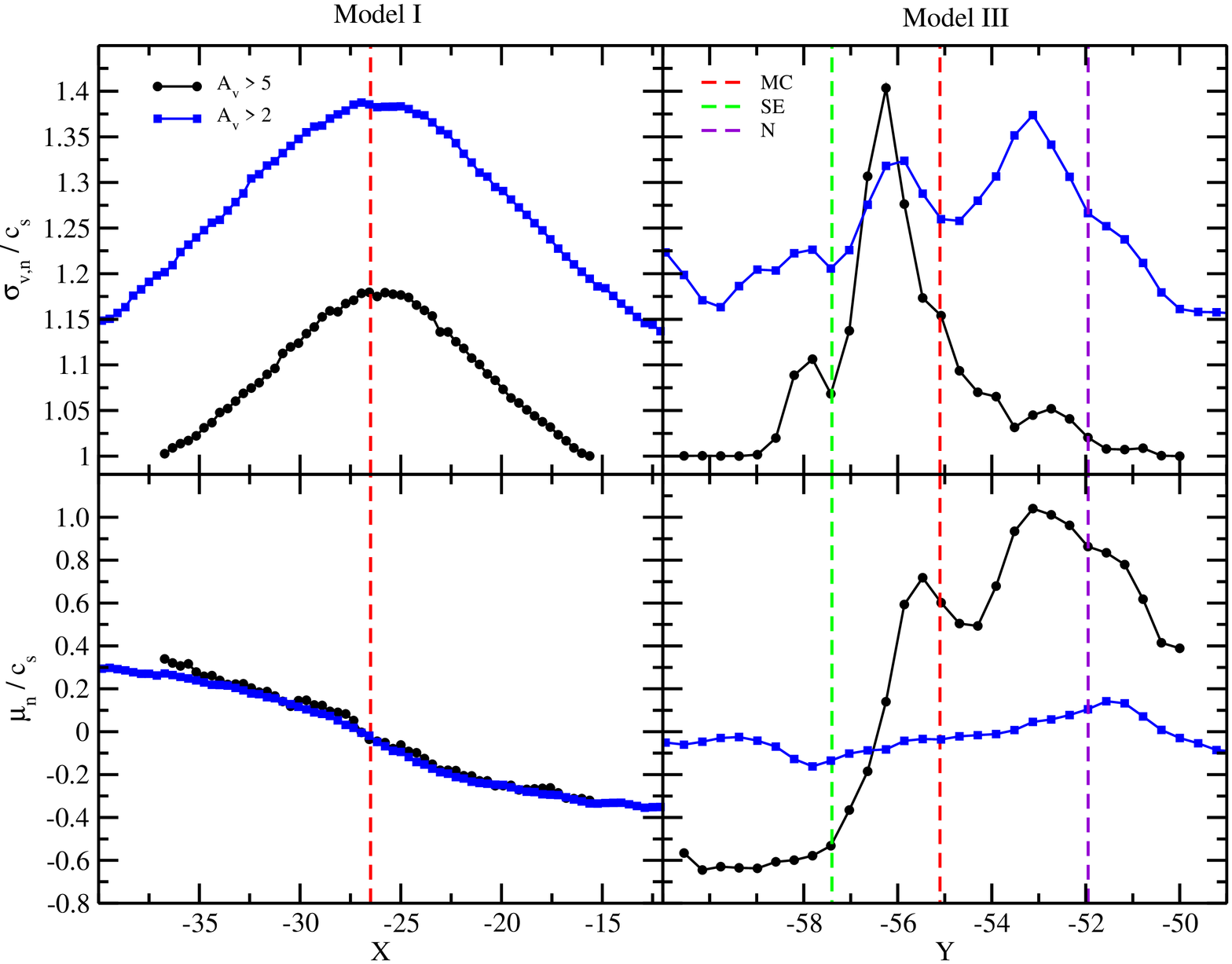}
\caption{Gaussian parameters as a function of position. Top row shows the variation in the standard deviation for neutral particles ($\sigma_{v,n}$) while bottom row shows the variation in the centroid velocity of the neutral particles ($\mu_{n}$). Both values are scaled to the sound speed ($c_{s}$). Left column shows the scans across the main core in Model I for NS lines of sight. Right column shows the scans across the full clump region in Model III for EW lines of sight. Blue lines shows the values for all gas with $A_{V}~>~2 $ mag while black lines show the values for all gas with $A_{V}~>~5$ mag. Vertical dashed lines indicate the locations of the main core (red) and SE core (green) for each of the models/lines of sight. The vertical purple line shows the location of a medium density region at $(x,y) = 14.8,~-51.95$ (north of the MC) denoted as N in the legend. }
\label{GaussianI}
\end{figure}

\subsubsection{Results: Region properties}

In addition to the individual lines of sight through the center of the cores, we performed scans across the clumps/cores for both the NS and EW lines of sight to see how the standard deviation or velocity dispersion ($\sigma_{v,n}$) and centroid velocity ($\mu_{n}$) of the neutral particles change as a function of position in step-like models only. For this analysis, we consider all material above 2 mag and above 5 mag to probe overall motions across the cloud. Figure~\ref{GaussianI} shows examples of some of the situations that can arise within forming cores. Here we show a scan across the clump/core in Model I along the NS los (left), and scans across the clump in Model III along the EW los (right). For both models, the top row shows the variation in the neutral velocity dispersion ($\sigma_{v,n}$) while the bottom row shows the variation in the neutral centroid velocity ($\mu_{n}$). The dashed lines indicate the locations of the main core (red) and SE core (green) for each of the models/lines of sight. The purple dashed line shows the location of a medium density region just north of the MC in Model III. The blue data points include all gas along the line of sight with $A_{V} > 2$ mag while the black data points include all gas with $A_{V} > 5$ mag.

The two plots for Model I depict a best-case scenario within our simulations. Looking at the change in variance across the clump, we see that the amount of dispersion increases as we get closer to the central core and then decreases again as we move away creating a Gaussian like profile. The degree of variance is larger for the $A_{V} > 2$ mag data than the $A_{V} > 5$ mag data. This shows that the majority of the dispersion comes from the low density gas which agrees with the trends shown in Figure~\ref{gaussianparams}. All across the core (defined by the $A_{V} > 5$ curve) there is a clear drop in velocity dispersion compared to the lower density material surrounding the core. This is reminiscent of the transition to coherence found toward low-mass dense cores \citep{Pineda2010Coherence}. At the location of the core itself, there is a very small local minimum indicating that the dispersion within the core is smaller than its immediate surroundings, 0.39\% and 0.36\% for the $A_{V} > 5$ mag and $A_{V} > 2$ mag data respectively.

Looking at the lower left panel, we see that the variation in the centroid velocity reveals that it decreases monotonically as one scans across the region. This decrease reveals that to the left of the clump/core, most of the material is moving in the positive direction while to the right, most of the material is moving in the negative direction. At the location of the core itself, there is a distinct drop as the direction of the dominating velocity changes signs. The decreasing trend observed in the centroid velocity is a result of the angle between the line of sight and the orientation of the clump. In this model, the NS and EW lines of sight are both approximately at a 45 degree angle to the major axis of the clump and therefore at a 45 degree angle to the in-flowing material. This results in material moving in the positive direction dominating the sample on one side of the core and material moving in the negative direction dominating on the other side. If the lines of sight were parallel and perpendicular to the clump major axis, the positive and negative velocity contributions would cancel each other out resulting in a flat line. Comparing the two data sets, we see there is a negligible difference between the $A_{V} > 2$ mag and $A_{V} > 5$ data mag for $\mu_{n}$.

The right hand column in Figure~\ref{GaussianI} shows the scans for Model III along the EW lines of sight. Compared to Model I, this model shows the effect of having multiple structures within a clump and along the line of sight. Looking at the upper right hand panel, we see that the trends in the two data sets are different. In the region between y = -56 and y = -57, we see a spike in the $A_{V} > 5$ mag data that is larger than in the $A_{V} > 2$ mag data. This would indicate that there is a large variance in the $A_{V} > 5$ mag gas within this region that is diluted by the low density gas. Likewise, between y = -55 and y = -52, there is a large peak in the $A_{V} > 2$ mag gas that is not as pronounced in the $A_{V} > 5$ mag gas. Looking at the scan of the central velocity, we see that the more complicated clump/core structure of this model no longer results in a monotonically decreasing trend. Again, as with the dispersion plot, we see large differences between the two data sets.

Comparing the data trends in Figure~\ref{GaussianI} to the density enhancement, velocity and momentum maps (see Figures~\ref{sigmaoverlays},~\ref{velocity},~\ref{p_sigmaoverlays}~\&~\ref{sigma_poverlays}), we can start to pick out some of the features visible in the maps. First, the location of the cores appear to coincide with either a valley or sharp gradient for both the  $\sigma_{n}/c_{s}$ and $\mu_{n}/c_{s}$ values. With respect to the variance ($\sigma_{n}$) this would once again indicate a transition to coherence. This is even evident at the location of the medium density region (N) at the location designated by the purple dashed line. Second, the bottom right hand panel has a sharp increase between y = -57 and y = -55 in the $A_{V} > 5$ mag gas that shows a distinct switch in the direction of velocity. This is consistent with the velocity and momentum maps of the region (see Figure~\ref{velocity}~\&~\ref{sigma_poverlays}) which show a velocity switch within this region.

\section{Discussion}

\subsection{Velocity Structures}

\begin{figure*}
\centering
\includegraphics[width = 0.33\textwidth]{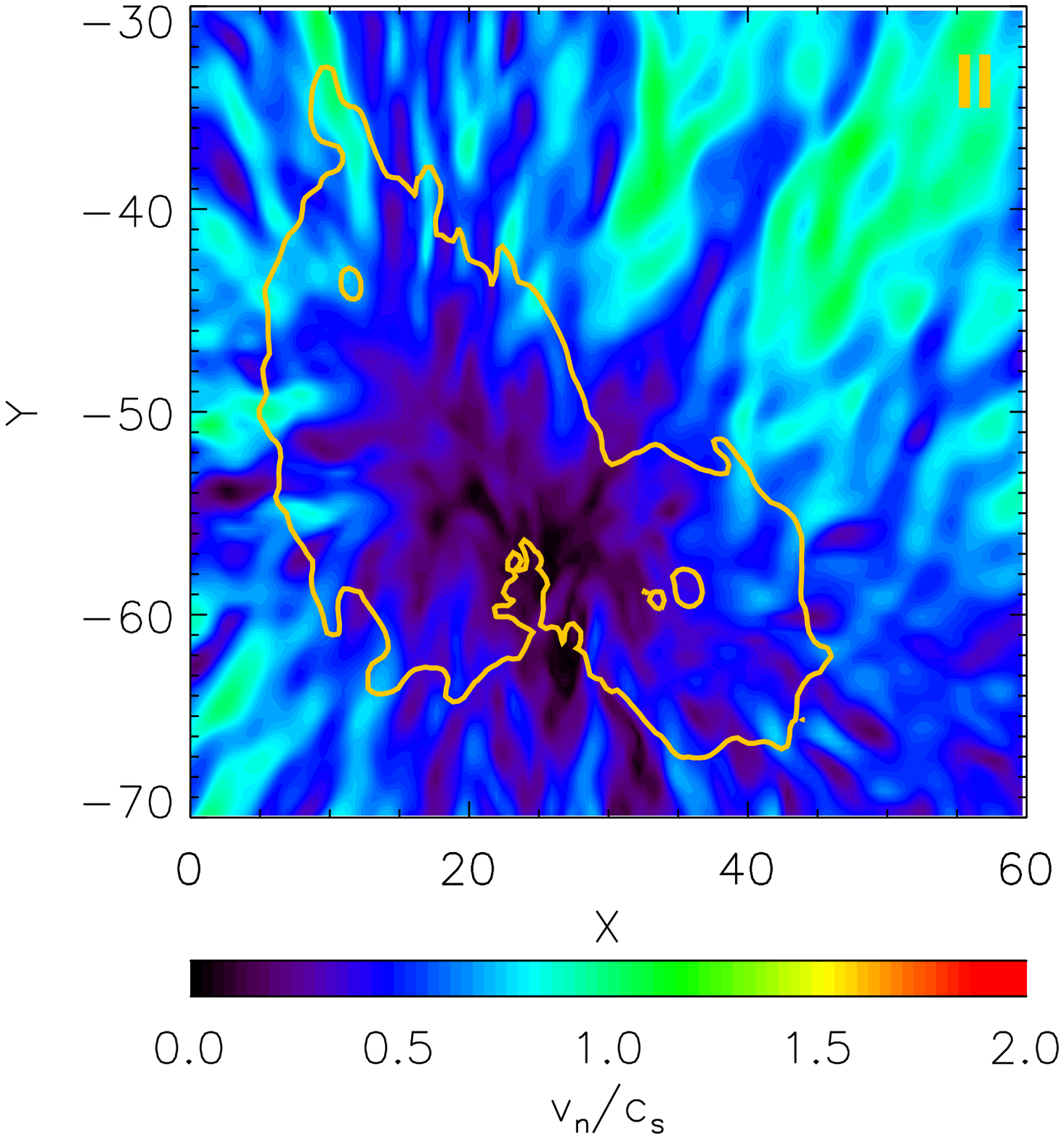}
\includegraphics[width = 0.33\textwidth]{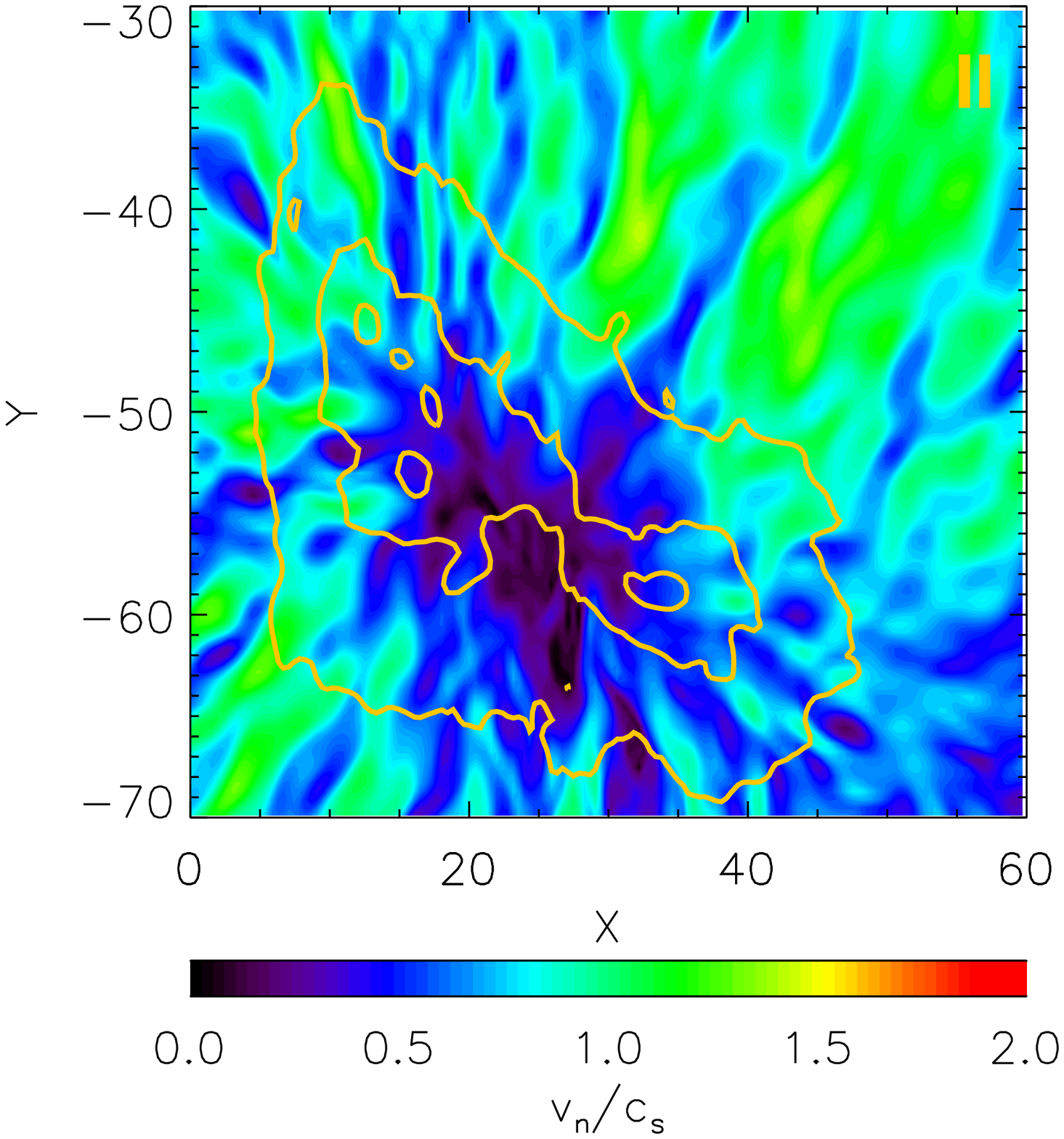}
\includegraphics[width = 0.33\textwidth]{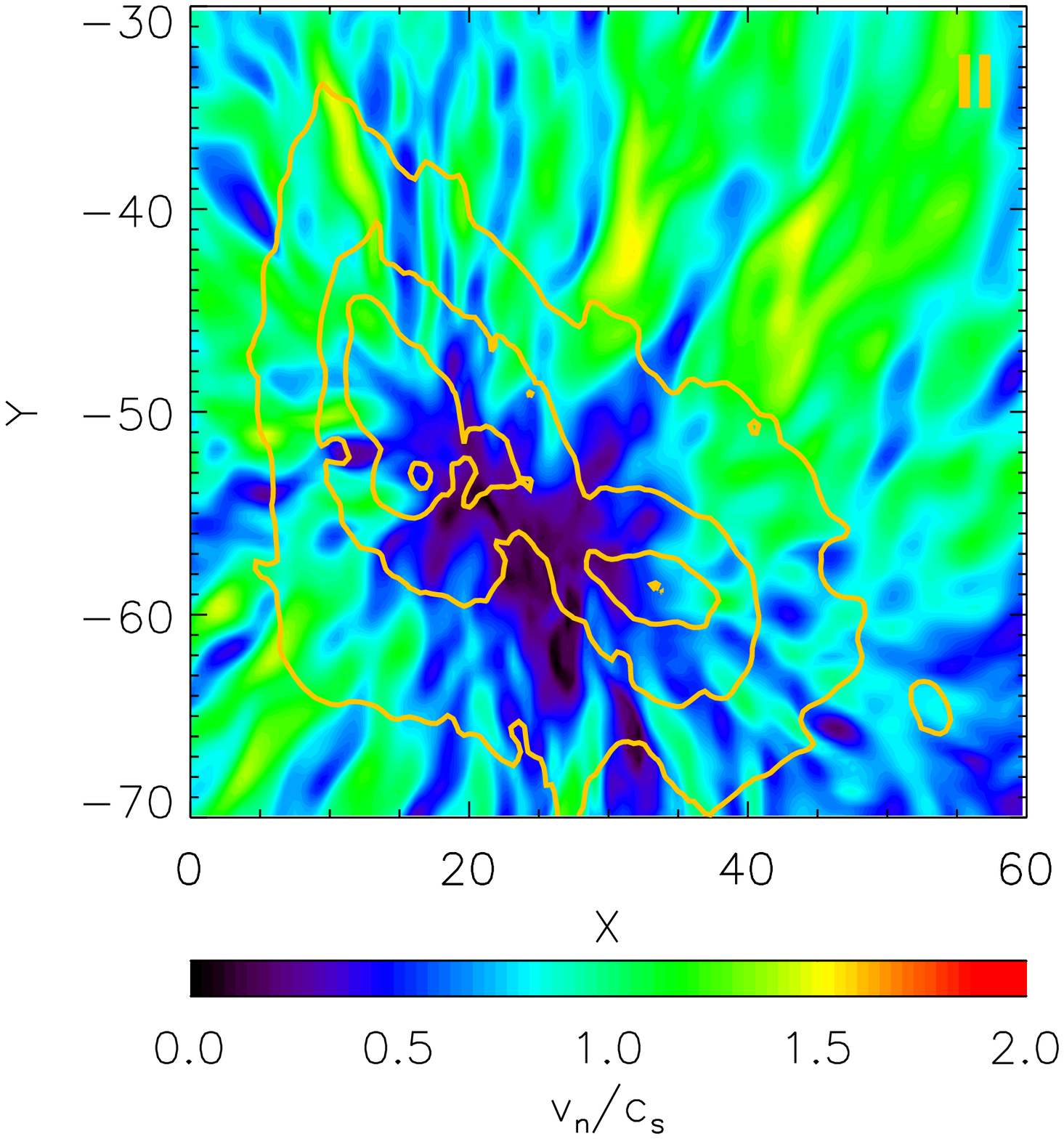}\\
\includegraphics[width = 0.33\textwidth]{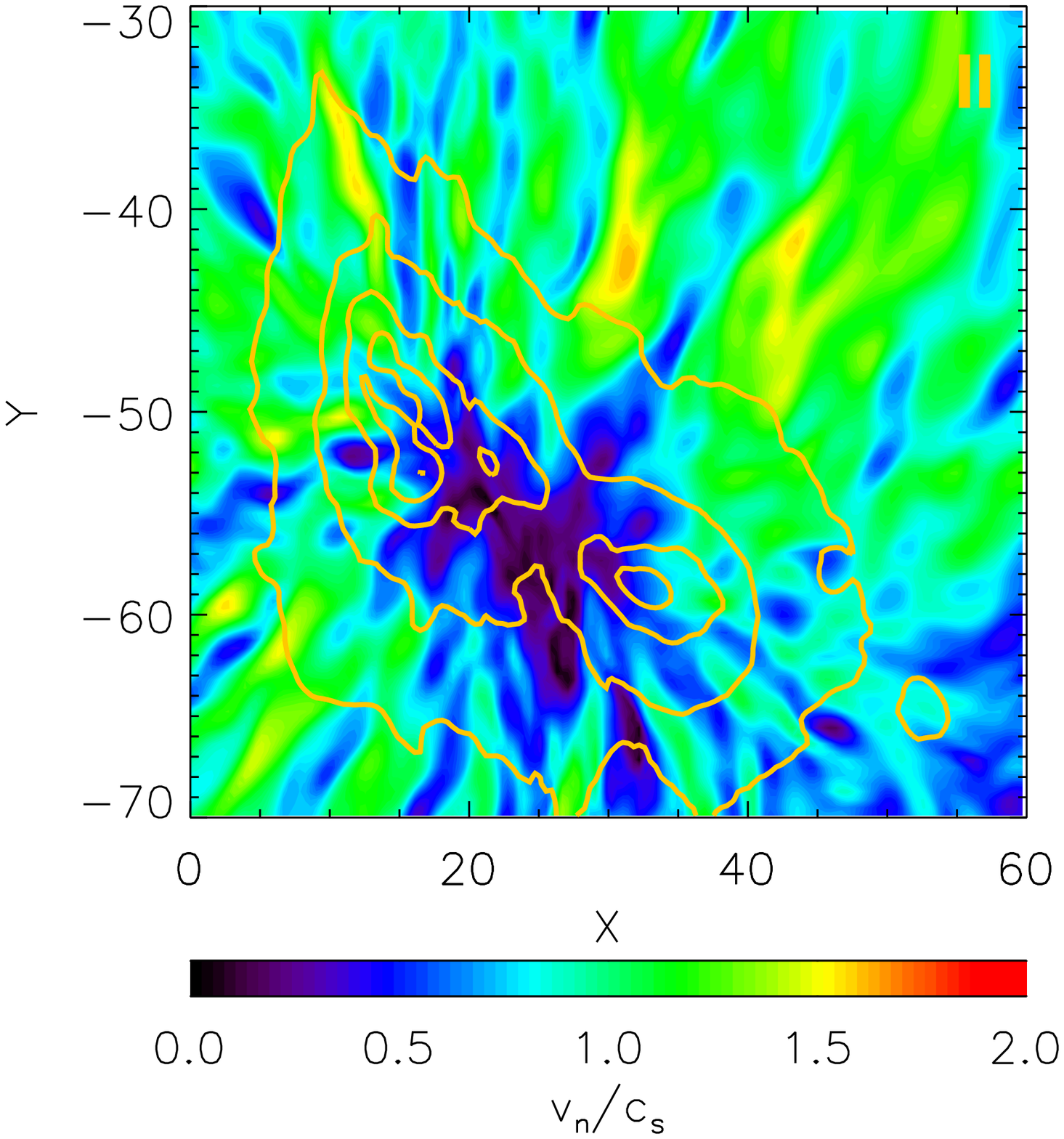}
\includegraphics[width = 0.33\textwidth]{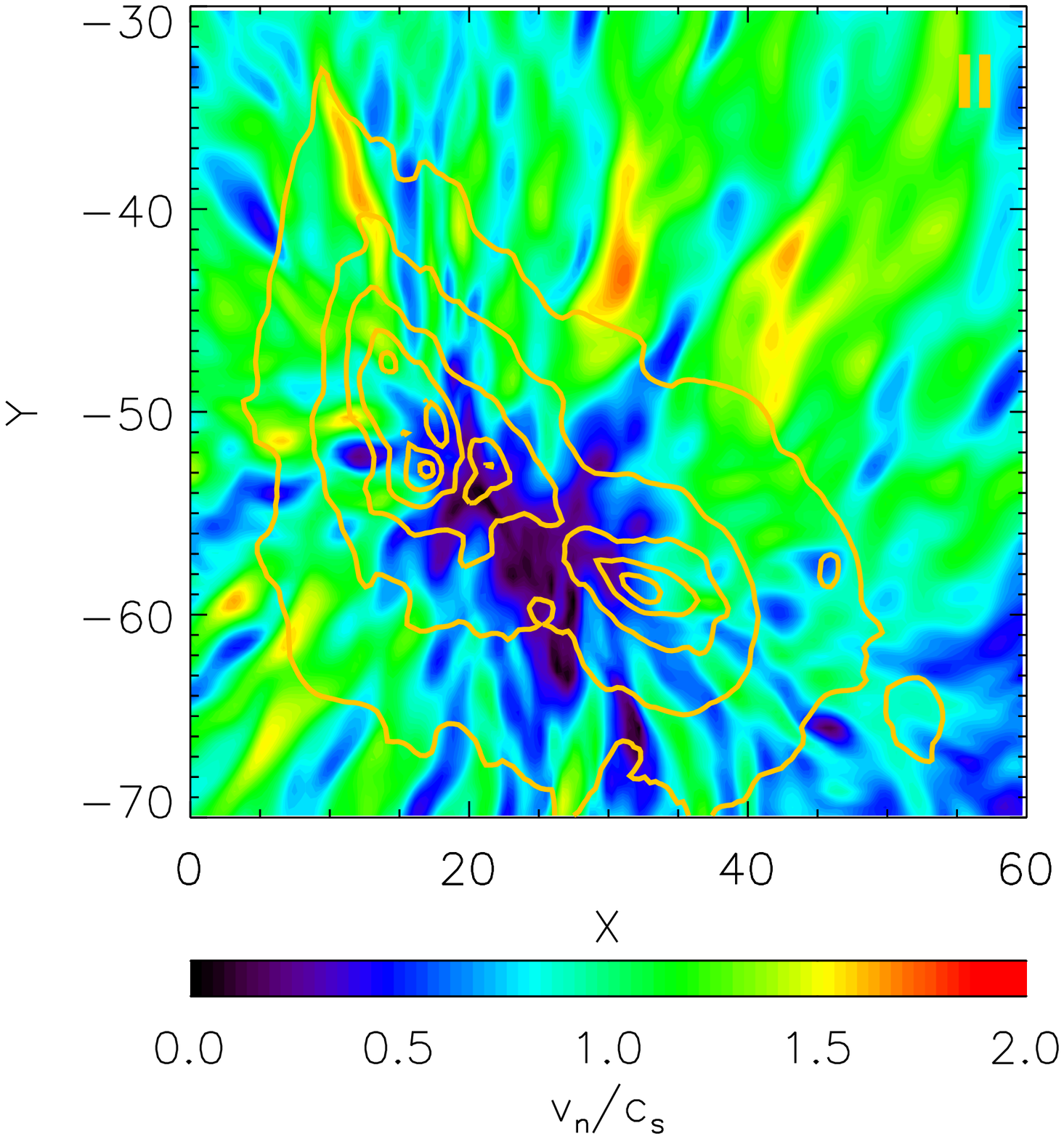}
\includegraphics[width = 0.33\textwidth]{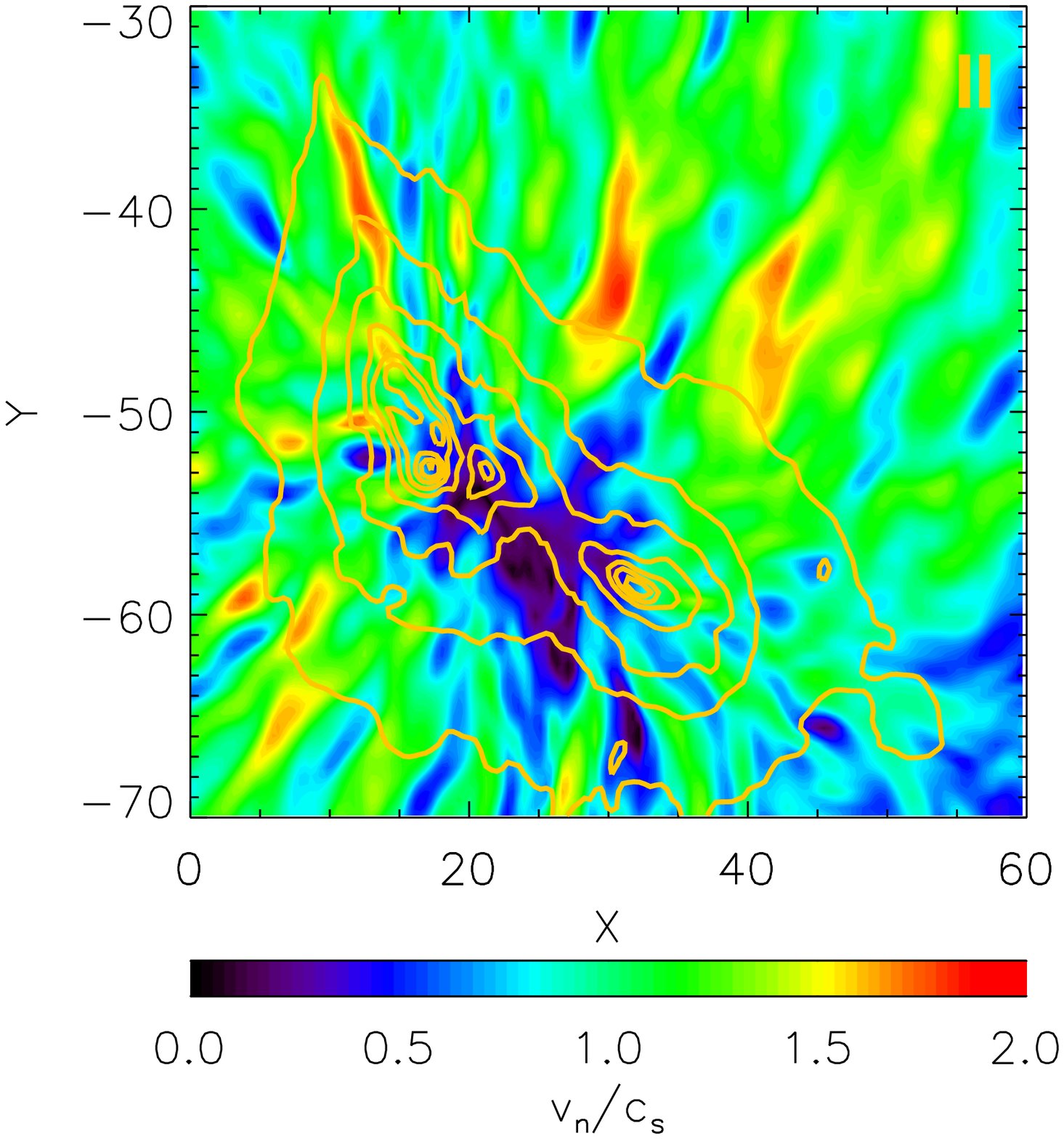}
\caption{Velocity maps for Model II for six times (in order from top left to bottom right: $t/t_{0}$ = 75.1, 78.1, 79.1, 79.5, 80.1 and 80.6). The contours show the visual extinction in 1 magnitude steps starting at $A_{V} = 2$ mag at the same time as the velocity map.}
\label{TimelapseII}
\end{figure*}

As shown in the velocity maps for Models II and III (Figure~\ref{velocity}, top middle and top right, respectively), we see that the cores that have formed within the clump seem to exist at the edge of the velocity valley. This is also observed in the velocity maps for Models IV - VI although to a lesser extent. Conversely, Model I shows that the position of the velocity valley coincides with the position of the core. The question is whether the cores always form within a velocity valley. Unlike observations, with simulations we have the advantage of being able to look back in time. Figure~\ref{TimelapseII} shows an example of the time-lapse of the velocity structure for Model II at times $t/t_{0} = 75.1, 78.1, 79.1, 79.6, 80.1~\&~80.7$ (from top left to bottom right, respectively). The contours show the density enhancement in 1 magnitude levels starting at $A_{V} = 2$ mag that correspond to the time of the velocity map. From these maps, we see that the extent of the velocity valley decreases over time. Looking at the visual extinction contours across the six times, the formation of the clump and subsequent formation of the cores at later times is evident. If we look at the location of the three high density regions in the last panel (Region 1: $x_{1},y_{1} = 17.6,-52.0$, Region 2: $x_{2},y_{2} = 32.0,-57.8$, and Region 3: $x_{3},y_{3} = 21.7, -52.8$) and compare to the previous times, we see that at early times, the regions which end up forming these three cores are initially entirely within the velocity valley and then migrate toward the edge of the velocity valley as they develop. This migration is due to the fact that the clump (and therefore the velocity valley) are contracting around these three regions. Comparing the final panel of Figure~\ref{TimelapseII} to the previous three panels, we can see that the positions of these three cores do not significantly change over the course of their formation (the magnitude of the core velocity is on the order of 0.07 pc/Myr). This implies that the cores do indeed seem to form within the velocity valley. This behaviour is also exhibited in the other 5 models (not shown). We also note that the cores are elongated along the flow with the densest regions occurring closer to the velocity valley. This results in an elongated shape with the head pointing toward the velocity valley. This is reminiscent to the cometary shapes of starless cores observed by \citet[][see Figure 14]{Crapsi2005}.

Based on the above, the following scenario regarding the formation of the clumps/cores emerges. As mentioned earlier, the components of the momentum reveal that in all models the clumps form where two flows converge (see Figure~\ref{sigma_poverlays}). This is consistent with the observations and simulation evidence of \citet{Schneider2010} and \citet{DB2011}. This convergence of flows creates the velocity valley (see Figure~\ref{velocity}). For the models which only form a single clump/core (Models I, IV, V and VI), the clump collapses and thus the velocity valley shrinks until a single core forms, causing the core to form directly over the velocity valley. This is due to the initial parameters of these models, specifically the lack of perturbations in Model I and the CR-only ionisation profile for Models IV - VI. Higher resolution simulations with box size reduced to $8\pi L_{0}$ (not shown) confirm that subfragmentation does not occur within the core for these models. For Models II and III, the scenario becomes more complicated since they both show evidence of the two stage fragmentation \citepalias[c.f.][]{BB2014}. For these two models the high ionisation fraction within the low density gas causes a parsec size clump to form around the initially large velocity valley.  As the clump collapses, the density increases and the velocity valley shrinks. This allows for a second fragmentation event to occur in regions where the density has risen enough to cause the ionisation fraction to drop, allowing for cores to form. Given the velocity structure, i.e. low velocity in the centre of the clump and higher velocity on the outskirts, the cores can only form from material flowing in from the outskirts of the clump. As the cores form from this in-flowing material, the clump continues to collapse on its own timescale \citep{BB2012}, resulting in cores coinciding with the periphery of the velocity valley.  Thus the velocity valley  may be at the origin of the ``transition to coherence'' widely observed in dense cores. For Models II and III, the in-flowing material can be mildly supersonic resulting in supersonic relative velocities between the two flows. This could result in shocks occurring at the junction point between the converging flows, however at the moment, the code assumes isothermality throughout the evolution.

\subsection{Synthetic Spectra}
\subsubsection{Comparison with Previous Work}
As mentioned in Section~\ref{aims}, star forming regions show three kinematic properties of starless cores in relation to their surroundings. First, cores are observed to have little internal turbulence \citep{BM1989,Jijina1999}, i.e., the velocity dispersion is dominated by thermal motions. This property is evident in all of our simulations as shown by Figure~\ref{gaussianparams}. Second, cores have smaller velocity dispersions than the surrounding material \citep{BM1989, Goodman1998, Jijina1999, Pineda2010Coherence}. Our simulations show evidence of this property as depicted by the dips in the $\sigma_{v}$ scans across cores (c.f Figure~\ref{GaussianI}), however contamination along the line of sight and small sampling of the core regions itself makes spotting such a dip non-trivial. A previous study of synthetic spectra by \citet{Kirk2009} was able to show that the cores in their simulations follow the first two properties (i.e., they had little internal turbulence and smaller velocity dispersions than the surrounding material) however they were not able to show evidence of the third (i.e., cores show small relative motions with respect to the surrounding material). As described by \citet{Walsh2004}, the diagnostic for determining the relative motion of the core to the surroundings is comparing the difference between the centroid velocity of the core and the low density gas to the linewidths of typical chemical tracers. Small relative motions are indicated by small differences in line center velocities similar to the line width of N$_{2}$H$^{+}$ while large relative motions would be indicated by shifts in centroid velocity comparable to the broader CO line widths. Simulations by \citet{Ayliffe2007} found regions  where the high density tracers would have larger line widths than the low density tracers, thus contradicting \citet{Walsh2004}, however they concede that by including chemical evolution to their calculations yields results that agree better with the observations.

The ionisation profile within our simulations depicts a rudimentary treatment of chemical evolution within the cloud. As discussed previously, we assume the neutral LDG component is CO while the ions in the cores correspond to N$_{2}$H$^{+}$. For our cores to have low relative motions compared to the surroundings, the spread in the difference between the core centroid velocity and the LDG centroid velocity must be smaller than the linewidth of the core gas. 
\begin{figure}
\includegraphics[width = 0.38\textwidth, angle = -90]{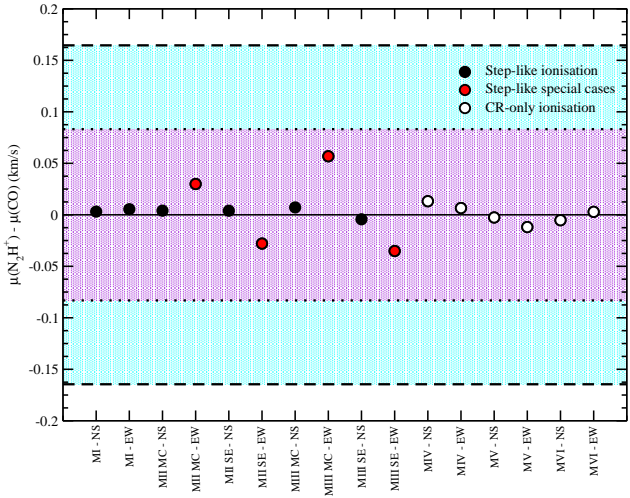}
\caption{Relative motion of core with respect to low density gas for all synthetic spectra. Filled regions between the dashed and dotted lines indicate area covered by the average FWHMs of the CO (blue + pink) and the N$_{2}$H$^{+}$ (pink only) linewidths, respectively. The solid line shows where zero deviation occurs and is plotted for visual reference. Solid black dots depict models with the step-like ionisation profile while open black dots depict models with the CR-only ionisation profile. Red dots indicate four cases with the step-like ionisation profile that exhibit larger deviations than the other spectra.}
\label{motions}
\end{figure}
Figure~\ref{motions} shows the comparison of the centroid velocity of the core minus the centroid velocity of the LDG ($\mu(N_{2}H^{+}) - \mu(CO)$) for all synthetic spectra. Solid black dots indicate models with the step-like ionisation profile while open black dots indicate models with the CR-only ionisation profile. The filled regions within the dashed and dotted lines indicate the extremes of the average CO (blue + pink) and N$_{2}$H$^{+}$ (pink only) linewidths, respectively. As one can see, the difference is relatively small for most cases and fall well within the boundaries defined by the N$_{2}$H$^{+}$ linewidth. This agrees with the observations of \citet{Walsh2004} as well as the concession for chemical evolution by \citet{Ayliffe2007}. 

The four red points in Figure~\ref{motions} indicate outliers that although they fall within the core linewidth boundaries, deviate greatly compared to the other models. Looking at the orientations of the los for these four points, we see that they all occur for EW lines of sight within Models II and III. Indeed, these lines of sight have shown the tendency to deviate compared to the other models in Figure~\ref{comparison1}~(right panel). Initial analysis of this phenomena assumed it was due to contamination by a second source along the line of sight, however this postulate is not consistent for all the lines of sight that show large deviations. For example, MII MC NS goes through two distinct high density regions and does not show a large deviation in $\mu(N_{2}H^{+}) - \mu(CO)$ while MII SE EW has very little contamination along the line of sight but does show a large deviation. Closer examination of the conditions for these four lines of sight reveal that they all intersect the imaginary line that defines where the direction of the velocity switches abruptly from positive to negative and the contributions from low and high density gas on either side of this line is asymmetric. This asymmetry is key to the large deviations in $\mu(N_{2}H^{+}) - \mu(CO)$. To illustrate, we compare Model I (which shows very little deviation) to Models II and III (which have evidence of significant deviation). In Model I, both lines of sight intersect the velocity switch line, however the centroid difference, as shown in Figure~\ref{motions}, is small. This is due to mostly symmetric contributions of low and high density gas moving in both directions along the line of sight which cancel each other out. The small deviation observed in Figure~\ref{motions} indicates a slight asymmetry between the motions of high density gas with respect to low density gas along the line of sight. Conversely, for Model II and Model III, the gas along the EW los is highly asymmetric about the velocity switch line causing the majority of the high density gas to be moving in one direction while the low density gas is moving in the opposite direction, resulting in a large deviation as shown in Figure~\ref{motions}. Given that the spectra are mass weighted, the sign of the deviation reveals which side of the velocity line the core is on: a positive deviation corresponds to being in the foreground along the line of sight moving away from the observer while a negative deviation corresponds to being in the background moving toward the observer. This explains why the phenomenon does not show up for MII MC NS. For this line of sight, both the cores and the cloud are moving in the same direction with very little deviation between the two. 

Observations of star forming regions (or any astronomical region) are two dimensional by nature with information about the third filled in by the velocity along the line of sight. By measuring the deviations in $\mu(N_{2}H^{+}) - \mu(CO)$ from zero as shown in Figure~\ref{motions}, one can start to construct a three dimensional picture of what a clump or other high density region may look like. In our simulations, we have the luxury of being able to see what the velocity and density look like in the same plane. For example, an observer of the region in Model III may consider the two clumps to be independent of each other, however by looking at the velocity map we can see that these two clumps are actually part of one larger clump since they share the same velocity valley. If they were independent, they would each have their own individual velocity valley. This type of analysis is not possible in observations. However, observations of a significant deviation in the mean velocities for the low density and high density gas along the line of sight have been reported \citep[e.g.,][]{Henshaw2013}. We interpret this as the presence of a velocity switch along the line of sight with the high density gas in the foreground.

As indicated above, observations of this phenomenon requires two very specific conditions. First the line of sight has to intersect the imaginary line where the velocity abruptly changes signs. Second, there needs to be asymmetry between the low- and high-density gas along the line of sight. Therefore, this large deviation in $\mu(N_{2}H^{+}) - \mu(CO)$ is not expected to be detected in all observations. \citet[][Figure 2, second panel]{Walsh2004} shows a small number of sources which have larger deviations in $\mu(N_{2}H^{+}) - \mu(CO)$ that still lie within the confines of the N$_{2}$H$^{+}$ line width, similar to those within this study. Our simulations are more appropriate for more quiescent low mass star forming regions, compared to those toward high-mass star forming regions \citep[e.g.][]{Henshaw2013} who found $\mu(N_{2}H^{+}) - \mu(CO) = 0.18\pm0.04$

\subsubsection{Effect of Ionisation Profile}

The analysis in the previous sections has shown that the ionisation profile has a great effect on the synthetic spectra. Figures~\ref{spectra}~-~\ref{comparison1} highlight the various difference that arise due to the different ionisation profiles. The most obvious difference is the broadening of the spectra of the low-density gas. As discussed above, this broadening is directly due to the flux frozen conditions set up by the ionisation fraction within the low-density gas. By inhibiting movement of neutrals across the magnetic field, the non-thermal component that is introduced by the microturbulent perturbations cannot dissipate as efficiently as in the CR only models. In addition, as shown by the left hand panel of Figure~\ref{comparison1}, the ionisation profile has an effect on the difference in dispersions for the neutral and ion components of the gas, with the step-like ionisation resulting in smaller differences within the cores and low density gas than observed for models with the CR only profile. This again would be a direct consequence of the nearly flux frozen conditions during the early stages of evolution for the models with the step-like ionisation profile. Frequent collisions between the neutrals and ions within the gas with high ionisation fraction would result in similar velocities for each species while the ability for neutrals to slip past neutrals at lower ionisation fractions allows the neutrals to obtain a larger velocity than the ions. For the CR only ionisation profiles, the neutrals are always able to slip past the ions and thus obtain a larger velocity even in the low density gas. 

Analysis of the velocity structure via Figure~\ref{velocity} revealed that the models with the step-like ionisation profile produce larger contraction velocities than the CR only models. This indicates that large velocities require the presence of a high ionisation fraction while the low ionisation fraction is more conducive for quiescent conditions. This is due to the fact that regions with high ionisation fractions have larger fragmentation lengthscales. These regions, therefore, amass material from further away than the lower ionised regions, resulting in larger velocities. The transition between high ionisation to low ionisation in the step-like models could explain the transition to coherence within cores as observed by \citet{Goodman1998} and \citet{Pineda2010Coherence}, among others. A closer look at the core region analysed in \citet{Pineda2010Coherence} shows a gradient in density between the core and surrounding gas, with the highest density occurring within the quiescent coherent region. If we assume that the ionisation fraction within the quiescent region is low while the surrounding region has a higher ionisation fraction then the transition to coherence is simply the transition from a UV dominated ionisation fraction to a lower CR dominated ionisation fraction and the transition to coherence could indicate the physical region within which the ionisation fraction drops. A study of the deuteration fraction \citep{Caselli2002b} within the core and surrounding high density gas could give an indication of the ionisation fractions within these two regions to confirm this theory. If this is the case, comparing the thickness of the transition zone to the column density maps will help constrain the parameters of the step-like ionisation profile. In the future, we plan to include a simple chemical network and implement a variation in the cosmic-ray ionisation flux across the cloud \citep[as in][]{Padovani2011}, to more accurately determine the ionisation fraction across the molecular cloud and study the effects on the dynamical evolution.

\section{Summary}

Based on the above analysis and discussion, there are several results which we summarise below.

\begin{itemize}

\item Analysis of the density and velocity structures for all models show that clumps form at the intersection of converging flows. These flows are formed due to gravitational instability rather than a pre-existing turbulent flow. Velocity data reveals a large low velocity region (velocity valley) that occurs at the center of the clump due to the convergence of the flows.

\item Cores form within the velocity valley. For the monolithic collapse (Model I) the center of the core coincides with the center of the velocity valley. For models with subfragmentation, the cores form on the periphery of the valley from material that is flowing into the valley. 

\item CR-only models exhibit subsonic to transonic contraction velocities while the step-like ionisation profile models exhibit supersonic contraction velocities. Regions with high ionisation fractions have larger fragmentation lengthscales and amass material from further away, resulting in larger velocities. Therefore observed high velocities require high ionisation fractions while the quiescent nature of cores requires lower ionisation profiles. The step-like ionisation profile naturally leads to supersonic in-fall speeds since it allows for steep gradients in the ionisation fraction to occur as the profile steps down from high to low ionisation. These steep gradients allow for velocity enhancements which are caused by flows of material from high to low ionisation fraction regions as the ability for neutrals to slip past the magnetic field lines increases. The observed transition to coherence could be the transition from high to low ionisation fractions. As such, it is very important to consider the influence of the ionisation fraction and chemistry on the evolution of molecular clouds. 

\item Analysis of the synthetic spectra show that the low density gas spectra have larger dispersions than the high density spectra for models with the step-like ionisation profile. In addition, the line widths are consistently larger for neutral particles than ions in both the low- and high-density gas. 

\item A comparison of the difference between centroid velocity for the core gas as traced by $N_{2}H^{+}$ and the low density gas as traced by CO to the linewidths of these two tracers was performed. Results show that the difference for all models was well within the limits defined by the linewidth of the $N_{2}H^{+}$ gas. This indicates that all cores show non-ballistic motions with respect to the surrounding gas which agrees with the observations presented by \citet{Walsh2004, Walsh2007}.

\item Large deviations in the difference between the centroid velocities of $N_{2}H^{+}$ and CO ($\mu_{N_{2}H^{+}}$ and $\mu_{CO}$, respectively) coincide with lines of sight that intersect the convergence of the two flows that make up the clump. Such a variation requires that the line of sight intersect the velocity switch and that the region have an asymmetry between the contributions of the low and high density gas along the line of sight. Instances of this within observations would give insight into the hidden structure along the line of sight.

\end{itemize}

\section*{Acknowledgments}
SB was supported by a Natural Science and Engineering Research Council (NSERC) Discovery Grant. The research leading to these results has received funding from the European Research Council under the European Union's Seventh Framework Programme (FP/2007-2013) / ERC Grant Agreement n. 320620-PALs

\bibliography{thesis}{}

\begin{thebibliography}{41}
\expandafter\ifx\csname natexlab\endcsname\relax\def\natexlab#1{#1}\fi

\bibitem[{{Ayliffe} {et~al.}(2007){Ayliffe}, {Langdon}, {Cohl}, \&
  {Bate}}]{Ayliffe2007}
{Ayliffe}, B.~A., {Langdon}, J.~C., {Cohl}, H.~S., \& {Bate}, M.~R. 2007,
  MNRAS, 374, 1198

\bibitem[{{Bailey} \& {Basu}(2012)}]{BB2012}
{Bailey}, N.~D. \& {Basu}, S. 2012, ApJ, 761, 67

\bibitem[{{Bailey} \& {Basu}(2014)}]{BB2014}
---. 2014, ApJ, 780, 40

\bibitem[{{Basu} \& {Ciolek}(2004)}]{BC2004}
{Basu}, S. \& {Ciolek}, G.~E. 2004, ApJL, 607, L39

\bibitem[{{Basu} {et~al.}(2009{\natexlab{a}}){Basu}, {Ciolek}, {Dapp}, \&
  {Wurster}}]{Basu2009a}
{Basu}, S., {Ciolek}, G.~E., {Dapp}, W.~B., \& {Wurster}, J.
  2009{\natexlab{a}}, NewA, 14, 483

\bibitem[{{Basu} {et~al.}(2009{\natexlab{b}}){Basu}, {Ciolek}, \&
  {Wurster}}]{Basu2009b}
{Basu}, S., {Ciolek}, G.~E., \& {Wurster}, J. 2009{\natexlab{b}}, NewA, 14, 221

\bibitem[{{Benson} \& {Myers}(1989)}]{BM1989}
{Benson}, P.~J. \& {Myers}, P.~C. 1989, ApJS, 71, 89

\bibitem[{{Caselli}(2002)}]{Caselli2002b}
{Caselli}, P. 2002, P\&SS, 50, 1133

\bibitem[{{Caselli} {et~al.}(2002){Caselli}, {Benson}, {Myers}, \&
  {Tafalla}}]{Caselli2002a}
{Caselli}, P., {Benson}, P.~J., {Myers}, P.~C., \& {Tafalla}, M. 2002, ApJ,
  572, 238

\bibitem[{{Caselli} {et~al.}(1998){Caselli}, {Walmsley}, {Terzieva}, \&
  {Herbst}}]{Caselli1998}
{Caselli}, P., {Walmsley}, C.~M., {Terzieva}, R., \& {Herbst}, E. 1998, ApJ,
  499, 234

\bibitem[{{Ciolek} \& {Basu}(2006)}]{CB2006}
{Ciolek}, G.~E. \& {Basu}, S. 2006, ApJ, 652, 442

\bibitem[{{Crapsi} {et~al.}(2005){Crapsi}, {Caselli}, {Walmsley}, {et~al.}
}]{Crapsi2005}
{Crapsi}, A., {Caselli}, P., {Walmsley}, C.~M., {et~al.} 
 2005, ApJ, 619, 379

\bibitem[{{Dale} \& {Bonnell}(2011)}]{DB2011}
{Dale}, J.~E. \& {Bonnell}, I. 2011, MNRAS, 414, 321

\bibitem[{{Friesen} {et~al.}(2010){Friesen}, {Di Francesco}, {Shimajiri}, \&
  {Takakuwa}}]{Friesen2010}
{Friesen}, R.~K., {Di Francesco}, J., {Shimajiri}, Y., \& {Takakuwa}, S. 2010,
  ApJ, 708, 1002

\bibitem[{{Goldsmith}(2001)}]{Goldsmith2001}
{Goldsmith}, P.~F. 2001, ApJ, 557, 736

\bibitem[{{Goodman} {et~al.}(1998){Goodman}, {Barranco}, {Wilner}, \&
  {Heyer}}]{Goodman1998}
{Goodman}, A.~A., {Barranco}, J.~A., {Wilner}, D.~J., \& {Heyer}, M.~H. 1998,
  ApJ, 504, 223

\bibitem[{{Goodman} {et~al.}(1993){Goodman}, {Benson}, {Fuller}, \&
  {Myers}}]{Goodman1993}
{Goodman}, A.~A., {Benson}, P.~J., {Fuller}, G.~A., \& {Myers}, P.~C. 1993,
  ApJ, 406, 528

\bibitem[{{Guelin} {et~al.}(1977){Guelin}, {Langer}, {Snell}, \&
  {Wootten}}]{Guelin1977}
{Guelin}, M., {Langer}, W.~D., {Snell}, R.~L., \& {Wootten}, H.~A. 1977, ApJL,
  217, L165

\bibitem[{{Hacar} {et~al.}(2013){Hacar}, {Tafalla}, {Kauffmann}, \&
  {Kov{\'a}cs}}]{Hacar2013}
{Hacar}, A., {Tafalla}, M., {Kauffmann}, J., \& {Kov{\'a}cs}, A. 2013, A\&A,
  554, A55

\bibitem[{{Hartmann} {et~al.}(2001){Hartmann}, {Ballesteros-Paredes}, \&
  {Bergin}}]{Hartmann2001}
{Hartmann}, L., {Ballesteros-Paredes}, J., \& {Bergin}, E.~A. 2001, ApJ, 562,
  852

\bibitem[{{Hartmann} {et~al.}(2012){Hartmann}, {Ballesteros-Paredes}, \&
  {Heitsch}}]{Hartmann2012}
{Hartmann}, L., {Ballesteros-Paredes}, J., \& {Heitsch}, F. 2012, MNRAS, 420,
  1457

\bibitem[{{Henshaw} {et~al.}(2014){Henshaw}, {Caselli}, {Fontani},
  {Jim{\'e}nez-Serra}, \& {Tan}}]{Henshaw2014}
{Henshaw}, J.~D., {Caselli}, P., {Fontani}, F., {Jim{\'e}nez-Serra}, I., \&
  {Tan}, J.~C. 2014, MNRAS, 440, 2860

\bibitem[{{Henshaw} {et~al.}(2013){Henshaw}, {Caselli}, {Fontani}, {et~al.}
}]{Henshaw2013}
{Henshaw}, J.~D., {Caselli}, P., {Fontani}, F., {et~al.} 
 2013, MNRAS, 428, 3425

\bibitem[{{Jijina} {et~al.}(1999){Jijina}, {Myers}, \& {Adams}}]{Jijina1999}
{Jijina}, J., {Myers}, P.~C., \& {Adams}, F.~C. 1999, ApJS, 125, 161

\bibitem[{{Kirk} {et~al.}(2009){Kirk}, {Johnstone}, \& {Basu}}]{Kirk2009}
{Kirk}, H., {Johnstone}, D., \& {Basu}, S. 2009, ApJ, 699, 1433

\bibitem[{{McDaniel} \& {Mason}(1973)}]{MM1973}
{McDaniel}, E.~W. \& {Mason}, E.~A. 1973, The Mobility and Diffusion of Ions in
  Gases (New York: Wiley)

\bibitem[{{Mouschovias} \& {Ciolek}(1999)}]{MC1999}
{Mouschovias}, T.~Ch. \& {Ciolek}, G.~E. 1999, in NATO ASIC Proc. 540: The
  Origin of Stars and Planetary Systems, ed. C.~J. {Lada} \& N.~D. {Kylafis},
  305

\bibitem[{{Mouschovias} {et~al.}(2006){Mouschovias}, {Tassis}, \&
  {Kunz}}]{Mouschovias2006}
{Mouschovias}, T.~Ch., {Tassis}, K., \& {Kunz}, M.~W. 2006, ApJ, 646, 1043

\bibitem[{{Myers} {et~al.}(1991){Myers}, {Ladd}, \& {Fuller}}]{Myers1991}
{Myers}, P.~C., {Ladd}, E.~F., \& {Fuller}, G.~A. 1991, ApJL, 372, L95

\bibitem[{{Padovani} \& {Galli}(2011)}]{Padovani2011}
{Padovani}, M. \& {Galli}, D. 2011, A\&A, 530, A109

\bibitem[{{Pineda} {et~al.}(2010{\natexlab{a}}){Pineda}, {Goodman}, {Arce}, {et~al.}
}]{Pineda2010Coherence}
{Pineda}, J.~E., {Goodman}, A.~A., {Arce}, H.~G., {et~al.} 
 2010{\natexlab{a}}, ApJL, 712,  L116

\bibitem[{{Pineda} {et~al.}(2010{\natexlab{b}}){Pineda}, {Goldsmith},
  {Chapman}, {et al.}
}]{Pineda2010}
{Pineda}, J.~L., {Goldsmith}, P.~F., {Chapman}, N., {et~al.}
 2010{\natexlab{b}}, ApJ, 721, 686


\bibitem[{{Ruffle} {et~al.}(1998){Ruffle}, {Hartquist}, {Rawlings}, \&
  {Williams}}]{Ruffle1998}
{Ruffle}, D.~P., {Hartquist}, T.~W., {Rawlings}, J.~M.~C., \& {Williams}, D.~A.
  1998, A\&A, 334, 678

\bibitem[{{Schneider} {et~al.}(2010){Schneider}, {Csengeri}, {Bontemps}, {et~al.}
}]{Schneider2010}
{Schneider}, N., {Csengeri}, T., {Bontemps}, S., {et~al.} 
2010, A\&A, 520, A49

\bibitem[{{Shu} {et~al.}(1987){Shu}, {Adams}, \& {Lizano}}]{Shu1987}
{Shu}, F.~H., {Adams}, F.~C., \& {Lizano}, S. 1987, ARA\&A, 25, 23

\bibitem[{{Tafalla} {et~al.}(2004){Tafalla}, {Myers}, {Caselli}, \&
  {Walmsley}}]{Tafalla2004}
{Tafalla}, M., {Myers}, P.~C., {Caselli}, P., \& {Walmsley}, C.~M. 2004, A\&A,
  416, 191

\bibitem[{{Tafalla} {et~al.}(2002){Tafalla}, {Myers}, {Caselli}, {Walmsley}, \&
  {Comito}}]{Tafalla2002}
{Tafalla}, M., {Myers}, P.~C., {Caselli}, P., {Walmsley}, C.~M., \& {Comito},
  C. 2002, ApJ, 569, 815

\bibitem[{{Walsh} {et~al.}(2004){Walsh}, {Myers}, \& {Burton}}]{Walsh2004}
{Walsh}, A.~J., {Myers}, P.~C., \& {Burton}, M.~G. 2004, ApJ, 614, 194

\bibitem[{{Walsh} {et~al.}(2007){Walsh}, {Myers}, {Di Francesco}, {et~al.}
}]{Walsh2007}
{Walsh}, A.~J., {Myers}, P.~C., {Di Francesco}, J., {et~al.} 
 2007, \apj, 655, 958

\bibitem[{{Williams} \& {McKee}(1997)}]{WM1997}
{Williams}, J.~P. \& {McKee}, C.~F. 1997, ApJ, 476, 166

\end{thebibliography}
\bibliographystyle{apj}

\end{document}